\documentclass[reprint,superscriptaddress,amsmath,amssymb,aps,prb]{revtex4-2}

\usepackage{color}
\usepackage{graphicx}
\usepackage{dcolumn}
\usepackage{bm}
\usepackage{hyperref}
\usepackage{braket}
\usepackage{mathtools}
\usepackage{float}
\usepackage{txfonts}

\begin{document}
\preprint{APS/123-QED}

\title{Activity-induced ferromagnetism in one-dimensional quantum many-body systems}

\author{Kazuaki Takasan}
\thanks{These two authors contributed equally to this work.}
\affiliation{Department of Physics, University of Tokyo, 7-3-1 Hongo, Bunkyo-ku, Tokyo 113-0033, Japan}

\author{Kyosuke Adachi}
\thanks{These two authors contributed equally to this work.}
\affiliation{Nonequilibrium Physics of Living Matter RIKEN Hakubi Research Team, RIKEN Center for Biosystems Dynamics Research, 2-2-3 Minatojima-minamimachi, Chuo-ku, Kobe 650-0047, Japan}
\affiliation{RIKEN Interdisciplinary Theoretical and Mathematical Sciences Program, 2-1 Hirosawa, Wako 351-0198, Japan}

\author{Kyogo Kawaguchi}
\affiliation{Nonequilibrium Physics of Living Matter RIKEN Hakubi Research Team, RIKEN Center for Biosystems Dynamics Research, 2-2-3 Minatojima-minamimachi, Chuo-ku, Kobe 650-0047, Japan}
\affiliation{RIKEN Cluster for Pioneering Research, 2-2-3 Minatojima-minamimachi, Chuo-ku, Kobe 650-0047, Japan}
\affiliation{Institute for Physics of Intelligence, The University of Tokyo, 7-3-1 Hongo, Bunkyo-ku, Tokyo 113-0033, Japan}
\affiliation{Department of Physics, University of Tokyo, 7-3-1 Hongo, Bunkyo-ku, Tokyo 113-0033, Japan}

\date{\today}

\begin{abstract}
We study a non-Hermitian quantum many-body model in one dimension analogous to the Vicsek model or active spin models, and investigate its quantum phase transitions. The model consists of two-component hard-core bosons with ferromagnetic interactions and activity, i.e., spin-dependent asymmetric hopping. Numerical results show the emergence of a ferromagnetic order induced by the activity, a quantum counterpart of flocking, that even survives in the absence of ferromagnetic interaction.
We confirm this phenomenon by proving that activity generally increases the ground state energies of the paramagnetic states, whereas the ground state energy of the ferromagnetic state does not change. By solving the two-particle case, we find that the effective alignment is caused by avoiding the bound state formation due to the non-Hermitian skin effect in the paramagnetic state. We employ a two-site mean-field theory based on the two-particle result and qualitatively reproduce the phase diagram. We further numerically study a variant of our model with the hard-core condition relaxed, and confirm the robustness of ferromagnetic order emerging due to activity. 
\end{abstract}

\maketitle 

\section{Introduction}
Active matter physics has engendered a wealth of insightful studies and fascinating phenomena, providing perspectives in understanding emergent properties in a range of complex systems. These studies have mainly been conducted within classical systems, featuring assemblies of self-propelled particles, each of which independently consumes energy to generate motion or other actions. Classical active matter systems, involving biomolecules~\cite{needleman2017active}, bacterial swarms~\cite{aranson2022bacterial}, and multicellular flows~\cite{henkes2020dense,kawaguchi2017topological} as well as non-biological materials~\cite{narayan2007long,bricard2015emergent}, have offered considerable insight into the behaviors that arise from non-equilibrium conditions~\cite{marchetti2013hydrodynamics,shankar2022topological}.
Despite the extensive exploration within this realm, active matter principles in quantum many-body systems remain largely unexplored.

As our understanding of active matter deepens, it is increasingly becoming of interest to investigate its potential implications in quantum systems. 
The recent development of controllable experimental settings in quantum many-body systems has opened up a broad range of physics fields, including the investigation of open quantum systems~\cite{Daley2014, Berman2012, schafer2020tools, Syassen2008, Barontini2013, Luschen2017, Tomita2017, Honda2023}.
Importing the active matter principles to these systems could potentially reveal new aspects of non-equilibrium behaviors at quantum scales.

With this motivation, we have previously introduced a model of quantum active matter~\cite{Adachi2022} and showed how the analog of the motility-induced phase separation can be realized as a quantum phase transition in a non-Hermitian quantum many-body setup. 
The quantum phase transitions observed in this model can be considered as a change in macroscopic behavior induced by biases in path ensembles in a classical model~\cite{tociu2019dissipation}.
Other works have investigated one-body properties of active quantum systems, such as non-Hermitian quantum walks \cite{yamagishi2023defining} and driven quantum particles~\cite{zheng2023quantum}. 

In addition to the general interest in studying phase transitions induced by non-Hermiticity, examining quantum models may be of interest due to the potential link to solvable models. For example, the asymmetric simple exclusion process has been analyzed using the Bethe ansatz due to its connection to the XXZ spin chain~\cite{gwa1992six,de2005bethe}. Another example is the one-dimensional (1D) Hubbard model, whose non-Hermitian extensions have also been previously considered and solved~\cite{fukui1998breakdown,uchino2012spin}.
Even the common tools used in the study of quantum mechanics, such as the uniqueness and other properties of the ground state in a certain class of models, or exact methods to solve fewer-body problems, may become useful in settling issues that typically arise in the simulations of classical active matter.
Connecting the two regimes may bring new insights through the exchange of methodologies, enhancing the understanding of nonequilibrium/non-Hermitian physics~\cite{Ashida2020}.

Here we introduce a model of 1D quantum active matter with ferromagnetic interactions to study how flocking can emerge in a quantum setup.
First, we show by exact diagonalization that the ferromagnetic phase can be induced by activity in the ground state of this model, even at the limit of no ferromagnetic alignment.
To understand the mechanism, we prove that the ferromagnetic state always has the lowest real eigenvalue when activity is finite and the transverse magnetic field is zero.
By analyzing the two-particle case, we also show explicitly that a two-particle bound state appears in the paramagnetic state, which causes the ferromagnetic state to be relatively more stable.
We further construct a mean-field theory that incorporates the bound state formation and demonstrate how it qualitatively reproduces the phase diagram of the original model.
Lastly, we investigate a soft-core variant of the model, which can be thought of as a two-lane extension, and show that the basic properties are preserved.

\section{Model}
\begin{figure}
    \centering
    \includegraphics[width=8cm]{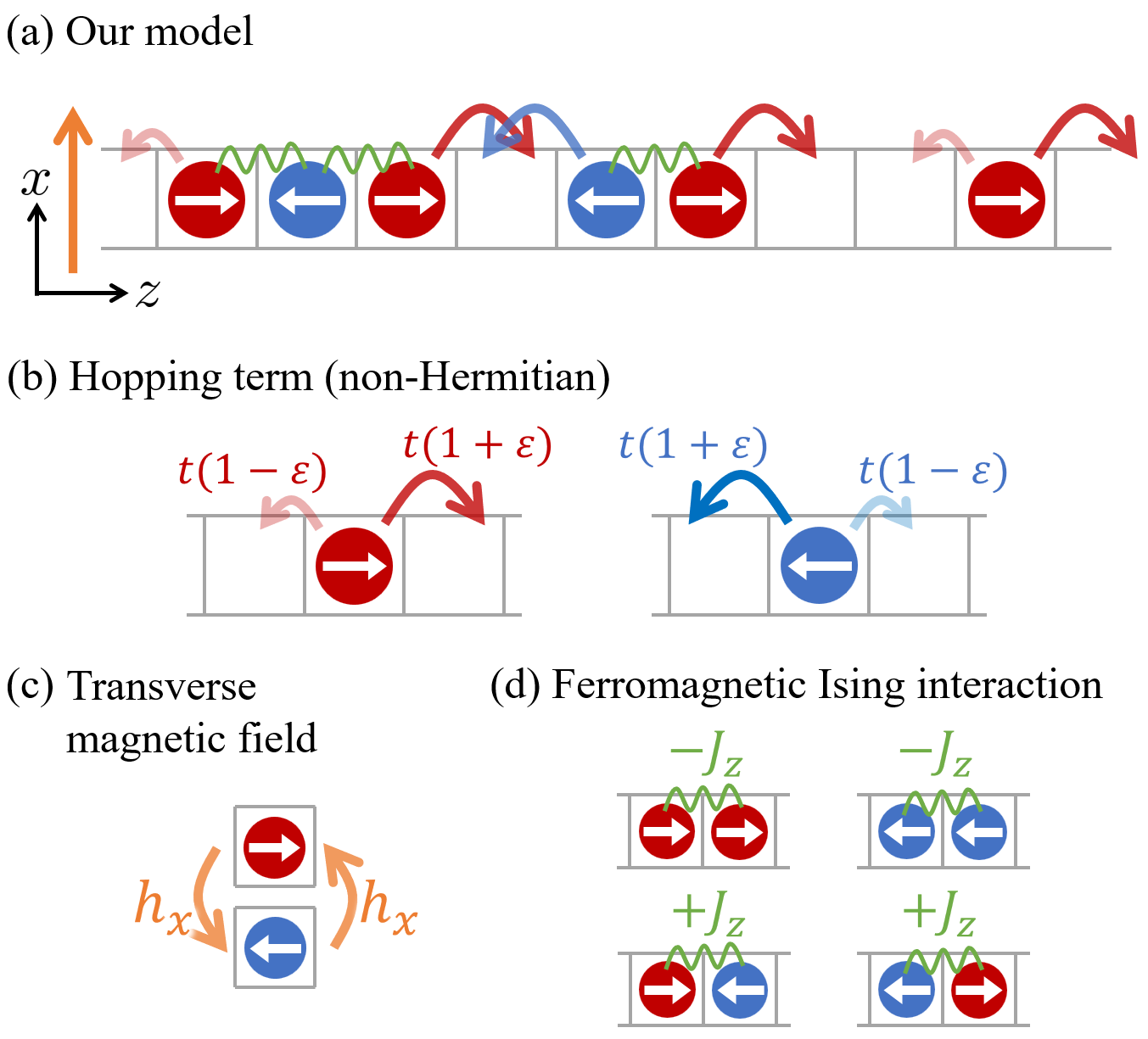}
    \caption{(a) The one-dimensional model \eqref{Eq:model_HCB1lane} for quantum active matter with aligning interaction. The chain is in the $z$-direction and the magnetic field (orange arrow) is applied in the $x$-direction. (b-d) Microscopic ingredients in the model. (b) corresponds to $\hat{H}_\mathrm{hop}$ [Eq.~\eqref{Eq:H_hop}] and $\hat{H}_\mathrm{act}$ [Eq.~\eqref{Eq:H_act}]. (c) and (d) are described by $\hat{H}_\mathrm{TFIM}$ [Eq.~\eqref{Eq:H_TFIM}].}
    \label{Fig:model}
\end{figure}
In this paper, we mainly consider non-Hermitian two-component (``spin-1/2") hard-core bosons in one dimension, which is schematically shown in Fig.~\ref{Fig:model}(a). The numbers of lattice sites and particles are denoted by $N$ and $L$, respectively. The particle density is given by $\rho := N/L$. The Hamiltonian is given as
\begin{equation}
\hat{H} := \hat{P}_{n \leq 1} \Big( \hat{H}_\mathrm{hop} + \hat{H}_\mathrm{act} + \hat{H}_\mathrm{TFIM} \Big) \hat{P}_{n \leq 1}, \label{Eq:model_HCB1lane}  
\end{equation}
where
\begin{gather}
    \hat{H}_\mathrm{hop} := -t \sum_{i=1}^L \sum_{s=\pm} (\hat{a}_{i+1,s}^\dag \hat{a}_{i,s} + \hat{a}_{i,s}^\dag \hat{a}_{i+1,s}), \label{Eq:H_hop} \\
    \hat{H}_\mathrm{act} := - \varepsilon t \sum_{i=1}^L \sum_{s=\pm} s~(\hat{a}_{i+1,s}^\dag \hat{a}_{i,s} - \hat{a}_{i,s}^\dag \hat{a}_{i+1,s}), \label{Eq:H_act} \\
    \hat{H}_\mathrm{TFIM} := - J_z \sum_{i=1}^L \hat{m}_i^z \hat{m}_{i+1}^z  - h_x  \sum_{i=1}^L \hat{m}_i^x. \label{Eq:H_TFIM}
\end{gather}
The parameters are taken as $t, J_z, h_x > 0$ and $0 \leq \varepsilon<1$. $\hat{a}^\dagger_{i, s}$ ($\hat{a}_{i, s}$) denotes the creation (annihilation) operator of a bosonic particle at the $i$th site ($i = 1, 2, \cdots, L$). Using these operators, the particle number operator is defined as $\hat{n}_{i,s}:=\hat{a}^\dagger_{i,s} \hat{a}_{i,s}$. Each component is labeled by $s \in \{ +, - \}$, which represents the spin direction along the $z$-axis, parallel to the 1D chain [see Fig.~\ref{Fig:model}(a)]. Unless otherwise specified, we consider the periodic boundary condition (PBC), i.e., $\hat{a}_{L+1, s}=\hat{a}_{1, s}$. The ``spin" operators are defined as $\hat{m}^z_i := \hat{n}_{i,+} - \hat{n}_{i,-}$ and $\hat{m}^x_i := \hat{a}^\dagger_{i,+} \hat{a}_{i,-} + \hat{a}^\dagger_{i,-} \hat{a}_{i,+}$. $\hat{P}_{n \leq 1}$ is a projection operator to the Hilbert space where the local particle number is not greater than one, which imposes the hard-core condition. $\hat{H}_\mathrm{hop}$ [Eq.~\eqref{Eq:H_hop}] is a usual hopping term between neighbouring sites. The non-Hermitian operator $\hat{H}_\mathrm{act}$ [Eq.~\eqref{Eq:H_act}] is an active hopping term that mimics the activity in classical active matter. This term consists of spin-dependent asymmetric hopping. Our previous work has shown that this term corresponds to the activity in classical active lattice gas models~\cite{Adachi2022}. $\hat{H}_\mathrm{TFIM}$ [Eq.~\eqref{Eq:H_TFIM}] is the Hamiltonian for the transverse-field Ising model (TFIM). We consider the ferromagnetic interaction ($J_z > 0$) that works similarly to the aligning interaction in classical active matter models~\cite{Vicsek1995, Solon2013}. The transverse field term enables the particles to change their direction by flipping the spin. At $\rho=1$, this model is reduced to the TFIM. The fermionic version of this model is the $t$-$J_z$ chain when the activity and magnetic field are both zero~\cite{Batista2000}. The fermionic Hubbard chain with non-Hermitian hopping ($\varepsilon > 0$) has been discussed in the context of spin-depairing transition in ultracold atoms~\cite{uchino2012spin}. This model is still exactly solvable using the Bethe ansatz even with the non-Hermitian term.

To study the quantum phase transition of this model, we investigate the ground state $\ket{\psi_\mathrm{GS}}$ of the Hamiltonian \eqref{Eq:model_HCB1lane}, which is defined as the energy eigenstate whose eigenvalue has the smallest real part~\cite{Adachi2022}. This is a convenient definition since the ground state energy will then become real and unique due to the Perron-Frobenius theorem, and the ground state corresponds to the steady state in a classical stochastic system with biases in path ensembles. This state is also relevant to experiment; starting from the Hermitian limit ($\varepsilon \to 0$), the system adiabatically evolves into the ground state~\cite{ashida2017parity} thanks to the uniqueness and the realness of the ground state. 

The active hopping term [Eq.~\eqref{Eq:H_act}] can be implemented by a single-particle loss~\cite{Adachi2022} induced by dissipative optical lattice, which has been recently realized to observe non-Hermitian skin effects~\cite{gou2020tunable,liang2022dynamic}. The effective time evolution by the non-Hermitian Hamiltonian can be obtained by neglecting the jump terms in the Lindblad quantum master equation, which is justified within a small time scale or when we do conduct post-selection to observe the quantum trajectory~\cite{Daley2014, gong2018topological}. Even without post-selection, the relaxation dynamics is expected to strongly reflect the nature of the effective non-Hermitian Hamiltonian since the eigenmodes of the corresponding Liouvillian containing only loss process are systematically constructed by the corresponding eigenstates~\cite{Torres2014}.

In the following, we denote the expectation value with respect to the ground state of the Hamiltonian \eqref{Eq:model_HCB1lane} as $\langle \cdots \rangle:= \bra{\psi_\mathrm{GS}} \cdots \ket{\psi_\mathrm{GS}}$ unless otherwise specified.

\begin{figure}[t]
    \centering
    \includegraphics[width=8.5cm]{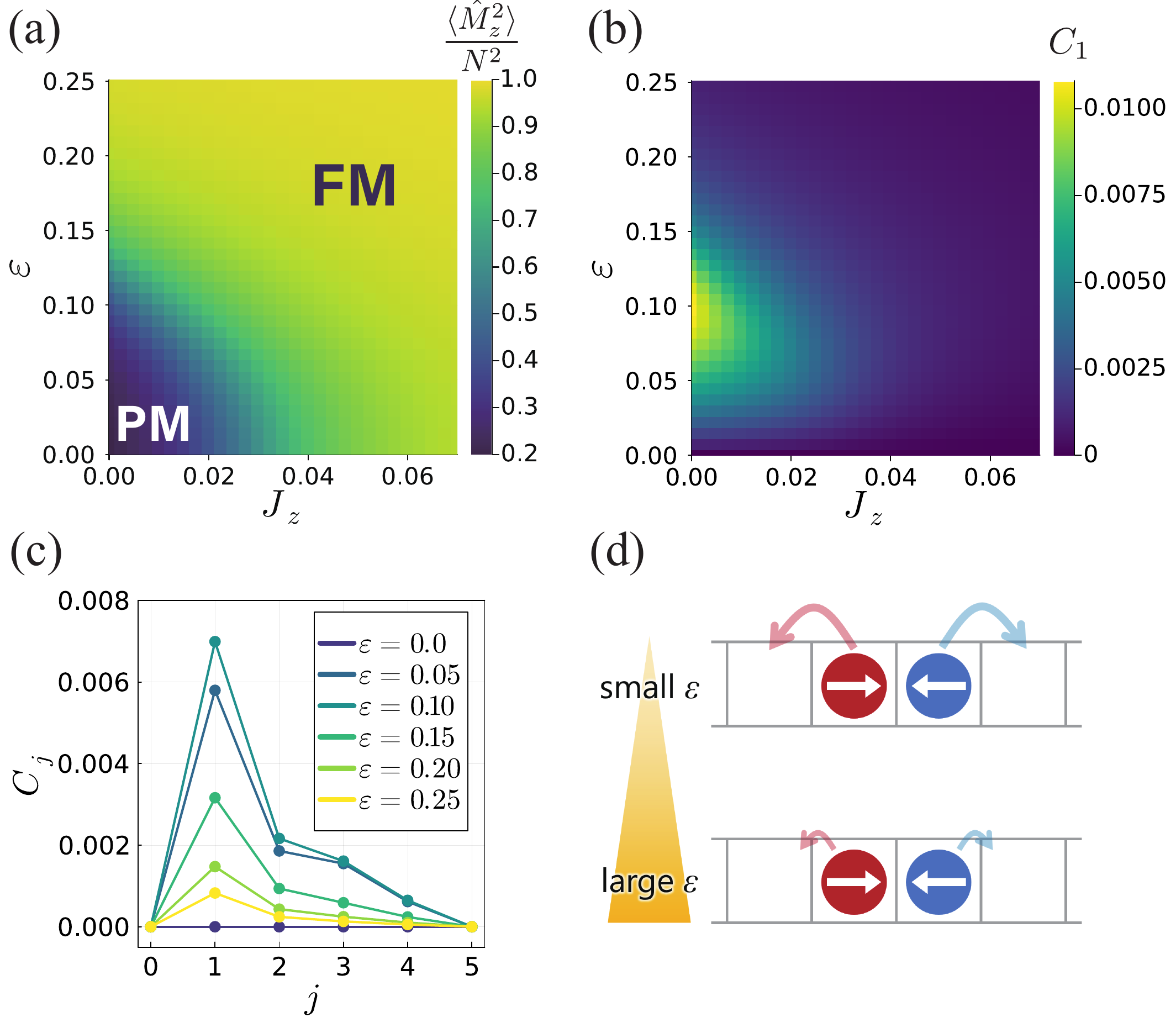}
    \caption{(a) Normalized squared magnetization $\langle \hat{M}_z^2 \rangle / N^2$, which is the order parameter for ferromagnetic (polar) order corresponding to a flocking phase. FM: ferromagnetic phase PM: paramagnetic phase. (b) Binding strength $C_1$, which takes a large value in the paramagnetic phase near the phase boundary. (c) Correlation function $C_j$ [$j=0, \cdots, 5 \, (=L/2)$] for $J_z=0.01$. (d) The $+-$ configuration behaves like a two-particle bound state under a large activity. The panels (a)-(c) are the numerical results based on the exact diagonalization of the Hamiltonian \eqref{Eq:model_HCB1lane} for $L=10$ and $\rho=0.5$, with $t=1$ and $h_x=0.01$. }
    \label{Fig:PD}
\end{figure}

\begin{figure}[t]
    \centering
    \includegraphics[width=8cm]{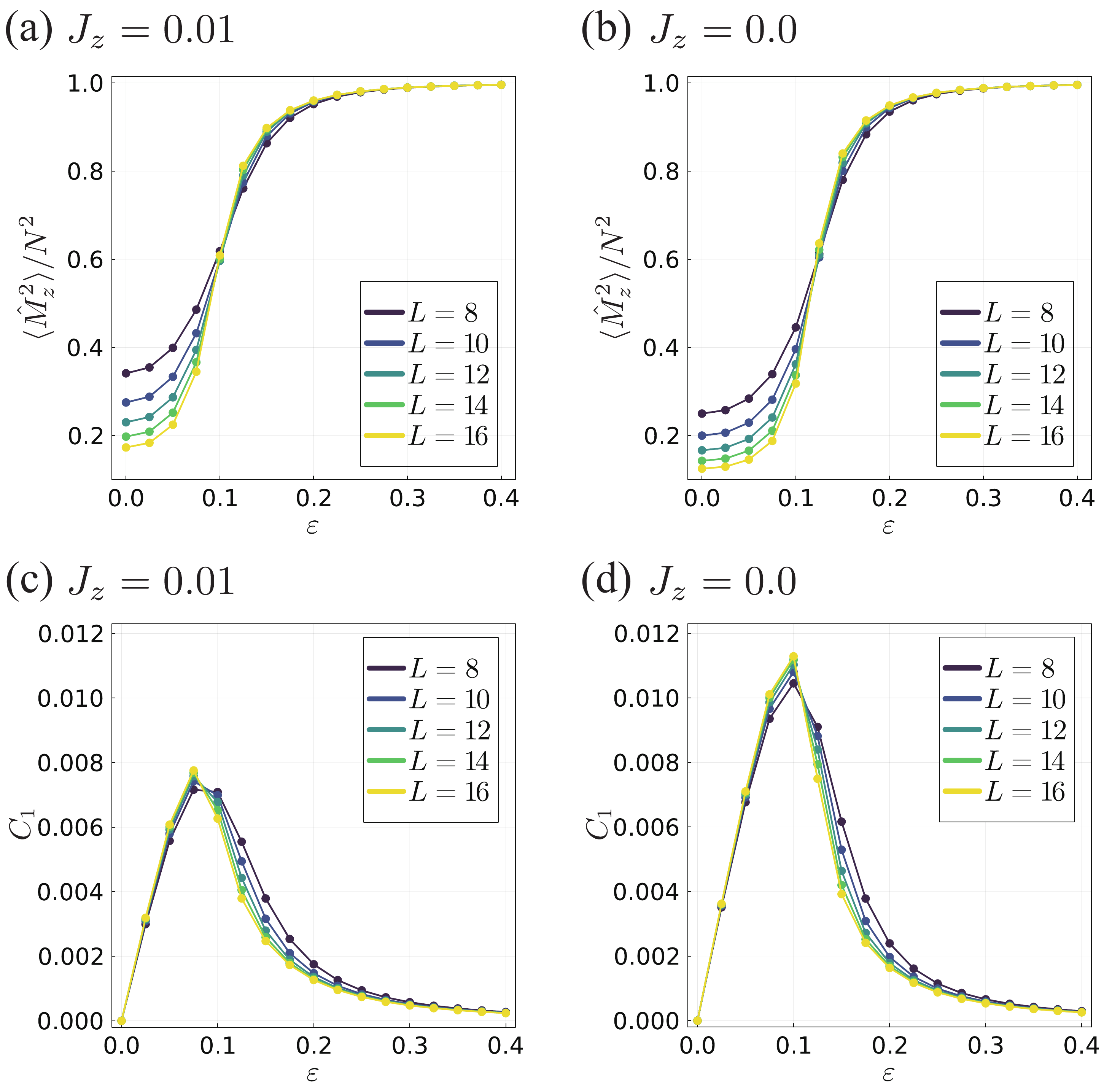}
    \caption{System-size dependence of (a,b) normalized squared magnetization $\langle \hat{M}_z^2 \rangle / N^2$ and (c,d) binding strength $C_1$. We used $J_z=0.01$ for (a) and (c) and $J_z = 0$ for (b) and (d). Here, all the numerical results are based on the exact diagonalization of the Hamiltonian \eqref{Eq:model_HCB1lane} for $L=8, \cdots, 16$ at $\rho=0.5$. We set the parameters as $t=1$, $h_x=0.01$.}
    \label{Fig:size_dep}
\end{figure}

\section{Numerical results}
\label{Sec:numerical_results}
In this section, we present numerical results based on the exact diagonalization of the Hamiltonian~\eqref{Eq:model_HCB1lane} for small system sizes, $L=8, \cdots, 16$. We find that (i) ferromagnetic order, a quantum counterpart of flocking, is induced by activity even without the aligning interaction, (ii) the activity-induced phase transition is captured by the correlation function [Eq.~\eqref{eq:Cij}] that captures a bound-state-like structure, and (iii) the ferromagnetic order is achieved with an infinitesimal activity for $h_x=+0$ and $J_z=0$. 

\subsection{Magnetization}
We examine the total magnetization $\hat{M}_z := \sum_{i=1}^L \hat{m}_i^z$, which is the order parameter for flocking since the spin degree of freedom corresponds to the dominant direction of hopping [see Figs.~\ref{Fig:model}(a) and (b)].
The calculated values of $\langle \hat{M}_z^2 \rangle / N^2 $ for different $\varepsilon$ and $J_z$ are shown in Fig.~\ref{Fig:PD}(a). For the finite activity regime, $\varepsilon > 0$, Fig.~\ref{Fig:PD}(a) shows that the parameter range of $J_z$ for the ferromagnetic ordered state is expanded by increasing the activity, meaning that activity enhances ferromagnetic order. Taking $J_z=0.01$, for example, the ground state exhibits a phase transition to the ferromagnetic state around $\varepsilon \sim 0.1$. By increasing the system size, the transition becomes sharper [Fig.~\ref{Fig:size_dep}(a)], which suggests that this activity-induced flocking survives even in the thermodynamic limit. 

Furthermore, the ferromagnetic order appears even without the ferromagnetic interaction, i.e., at $J_z=0$ [Fig.~\ref{Fig:PD}(a)], meaning that the quantum flocking transition can occur without the aligning interaction, which is in contrast to the flocking transitions in classical active matter that typically require aligning interactions.
This behavior can be related to the fact that our model is in the deeply quantum regime, as the model \eqref{Eq:model_HCB1lane} is far from the classical condition, where the non-Hermitian Schr\"odinger equation can be exactly mapped to the classical Markov process~\cite{Adachi2022}. As shown in Fig.~\ref{Fig:size_dep}(b), this transition also becomes steeper with increasing the system size, and thus is expected to appear even in the thermodynamic limit.

\begin{figure}[t]
    \centering
    \includegraphics[width=8.5cm]{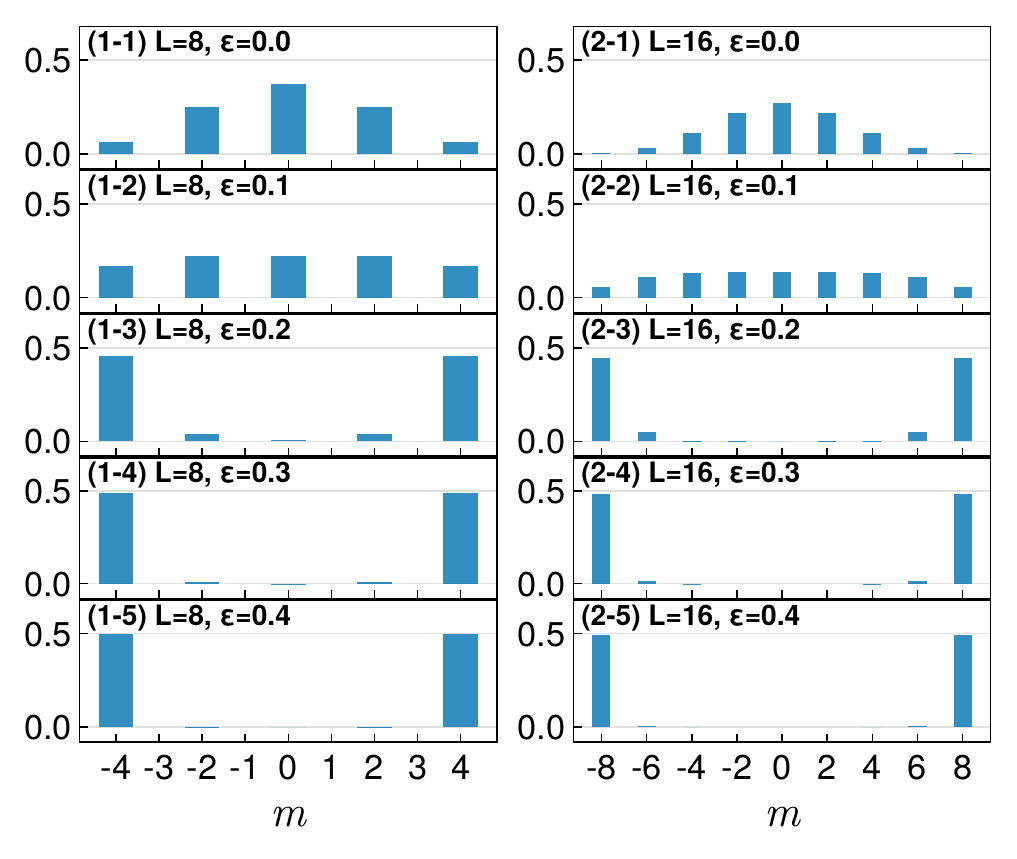}
    \caption{Distribution of the total magnetization $P_m$ for the ground state of the Hamiltonian~\eqref{Eq:model_HCB1lane} with $L=8, 16$ and $\varepsilon=0.0$-$0.4$. Other parameters are set as $\rho=0.5$, $t=1.0$, $J_z=0.0$, and $h_x=0.01$. }
    \label{fig:mdist}
\end{figure}

To further examine the difference from classical flocking transitions, we consider the distribution of the total magnetization $P_m$ for the ground state $\ket{\psi_\mathrm{GS}}$.
$P_m$ is represented as $P_m = || \hat{P}_m \ket{\psi_\mathrm{GS}} ||^2$ where $\hat{P}_m$ is the projection operator on the eigenspace of $\hat{M}_z$ with the eigenvalue $m$.
In Fig.~\ref{fig:mdist}, we show $P_m$ for several values of $\varepsilon$ with $L = 8$ or $16$.
While $P_m$ exhibits a broad peak around $m=0$ for $\varepsilon=0$, the sharp peaks are formed around $m=\pm N$ and the fraction at $m=0$ decreases to zero when increasing the activity $\varepsilon$.
This is in contrast to the flocking phase in the classical active Ising model~\cite{Solon2013}, where a finite fraction of zero magnetization is observed even in the large system size simulation.

\subsection{Correlation function and binding strength}

Figure~\ref{Fig:PD}(a) shows that the transition line of the activity-induced phase transition is smoothly connected to the transition point in the Hermitian case ($\varepsilon=0$). It is natural to ask whether the activity-induced phase transition can be distinguished from the ferromagnetic transition within the Hermitian system presented in this section. To answer this, we introduce the following correlation function,
\begin{equation}
    C_{ij} := \langle \hat{n}_{i, +} \hat{n}_{i+j, -} - \hat{n}_{i, -} \hat{n}_{i+j, +}  \rangle. \label{eq:Cij}  
\end{equation}
In practice, we use $C_j:=C_{1j}$ instead because $C_{ij}$ is independent of $i$ under the PBC. 

The typical behavior of $C_j$ is shown in Fig.~\ref{Fig:PD}(c); a peak appears at $j=1$ whose height increases approaching the transition point, and then decreases with increasing activity. This implies that $C_1$ works as a good indicator to detect the transition point of the activity-induced phase transition. Indeed, Fig.~\ref{Fig:PD}(c) shows that $C_1$ takes large values near the transition point on the $\varepsilon$-axis. On the other hand, this enhancement does not occur near $\varepsilon=0$. Therefore, the behavior of $C_1$ within the paramagnetic side distinguishes the activity-induced phase transition from the Hermitian transition triggered by $J_z$.

The enhancement of $C_1$ near the transition point reflects the appearance of a bound-state-like structure in the ground state $\ket{\psi_\mathrm{GS}}$. In the paramagnetic phase, both ~$+$ and~$-$ appear at an almost equal probability. First, let us consider the $+-$ configuration, where the~$+$ particle sits on the left adjacent site of the~$-$ particle, as shown in Fig.~\ref{Fig:PD}(d). With stronger activity, this configuration becomes more difficult to dissolve because the right (left) hopping of~$+$($-$) particle is prohibited due to the hard-core condition, and the hopping amplitude in the other direction is small [see Fig.~\ref{Fig:PD}(d)]. In contrast, the $-+$ configuration is more likely to be dissociated upon increasing the activity. 

The $+-$ configuration can be regarded as a bound state. Indeed, it can be shown that the two-particle wavefunction is exponentially localized in the relative coordinate for the $h_x \to 0 $ limit at $J_z=0$, which will be discussed further in Sec.~\ref{Sec:two_particle_problem}. From the definition of $C_{ij}$ [Eq.~\eqref{eq:Cij}], it is clear that $C_{1}$ measures the asymmetry between the bound ($+-$) and non-bound ($-+$) configurations and $C_{1}$ becomes larger when the bound configuration is dominant; we therefore call $C_1$ the \textit{binding strength}. The numerical results show that the binding strength is small in the ferromagnetic phase. This is reasonable since the balance between the number of $+$ and $-$ particles in the ferromagnetic phase becomes significantly biased and the bound pairs almost disappear.

Figures~\ref{Fig:size_dep}(c) and (d) show that the enhancement of binding strength near the ferromagnetic transition is robust even for a larger system. This suggests that the enhancement of binding strength can be used to capture the activity-induced phase transition. $C_{ij}$ is calculable from the spin-resolved local density $\langle \hat{n}_{i,s} \rangle$, which is measurable in ultracold atom experiments, e.g., by quantum gas microscope~\cite{schafer2020tools, Gross2017, Gross2021}.

\begin{figure}[t]
    \centering
    \includegraphics[width=8cm]{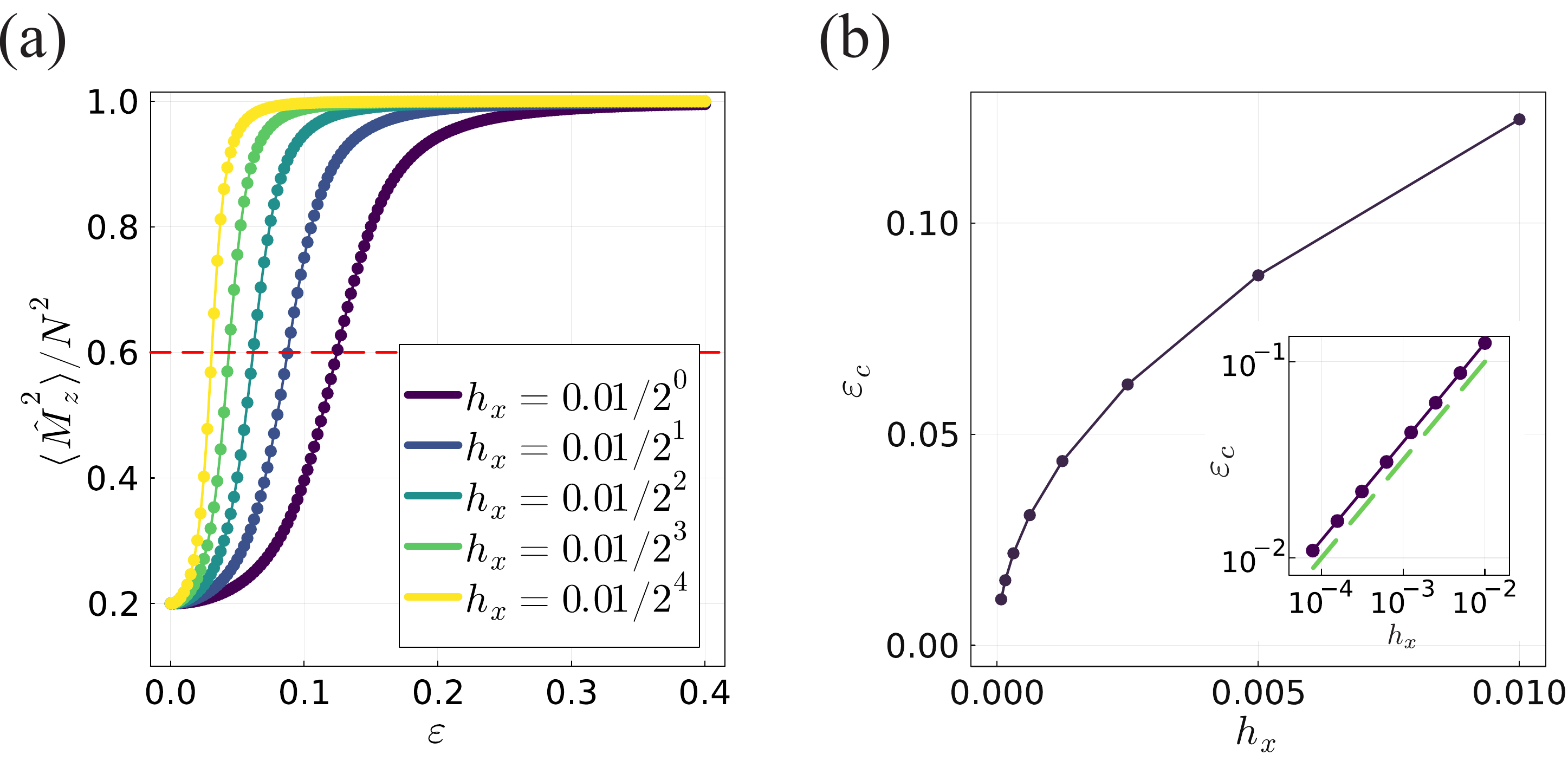}
    \caption{(a) $h_x$-dependence of normalized squared magnetization $\langle \hat{M}_z^2 \rangle / N^2$. (b) Transition point $\varepsilon_c$ for different values of $h_x$. Here, $\varepsilon_c$ is defined as the value of $\varepsilon$ where $\langle \hat{M}_z^2 \rangle / N^2$ takes 0.6, represented by a red dashed line in (a). The inset of the panel (b) is the log-log plot. The green dashed line represents $\varepsilon_c = h_x^{1/2}$ for comparison. All numerical results are based on the exact diagonalization of the Hamiltonian \eqref{Eq:model_HCB1lane} for $L=10$ at $\rho=0.5$. We set the parameters as $t=1$, $h_x=0.01$.}
    \label{Fig:h_dep}
\end{figure}

\subsection{$h_x$-dependence at $J_z=0$}
To clarify the nature of the activity-induced phase transition at $J_z=0$, we study its behavior with changing $h_x$. Figure~\ref{Fig:h_dep}(a) shows that the ferromagnetic order is induced by a smaller activity by decreasing $h_x$. This is reasonable since $h_x$ introduces quantum fluctuation between the $+$ and $-$ states and makes the paramagnetic phase favorable, as in the TFIM. 

To understand this behavior more quantitatively, we calculate the $h_x$-dependence of the transition point $\varepsilon_c$ [Fig.~\ref{Fig:h_dep}(b)]. We define $\varepsilon_c$ as the value of $\varepsilon$ where the normalized squared magnetization $\langle \hat{M}_z^2 \rangle/N^2$ takes 0.6 for convenience. We find $\varepsilon_c \sim {h_x}^{1/2}$, which indicates $\varepsilon_c \to 0$ in the limit of $h_x \to 0$. This means that the activity-induced phase transition occurs with an infinitesimal activity for $h_x=+0$ and $J_z=0$, allowing us to analyze the phase properties under the simple condition, $J_z=h_x=0$.

\section{Proof of the ferromagnetic ground state}
\label{Sec:proof}

Focusing on the case with no explicit ferromagnetic interaction (i.e., $J_z = 0$), we consider the mechanism of the activity-induced ferromagnetism observed in Sec.~\ref{Sec:numerical_results}.
Since the numerical results suggest $\varepsilon_c \to 0$ for $h_x \to 0$ (Fig.~\ref{Fig:h_dep}), we further set $h_x = 0$ and examine how nonzero $\varepsilon$ can stabilize ferromagnetism in the ground state.

Since the Hamiltonian $\hat{H}$ commutes with the total magnetization $\hat{M}_z$ for $h_x = 0$, the eigenvalue of $\hat{M}_z$, $M_z$, is a good quantum number.
In addition, since the positions of neighboring particles with different spins cannot be exchanged in the model~\eqref{Eq:model_HCB1lane}, the spin configuration of the particles except empty sites with the PBC taken into account can be specified by another quantum number, $S$.

For a given set of $(L, N, M_z)$, we define $S$ as follows.
First, we take a Fock state $\ket{g_1} := \ket{\underbrace{+ \cdots +}_{(N + M_z) / 2} \underbrace{- \cdots -}_{(N - M_z) / 2} 0 \cdots 0}$, where $\pm$ and $0$ represent the particle's spin state and empty site, respectively.
Then, we consider a partial Fock space $\mathcal{H}_1$ spanned by $\{ \text{Fock basis } \ket{f} | \exists n \in \mathbb{N} \text{ s.t. } \braket{f | \hat{H}^n | g_1} \neq 0 \}$, i.e., all the states that can be visited starting from $\ket{g_1}$ just by hopping and without flipping the spins.
Similarly, we take another Fock state $\ket{g_2} := \ket{\underbrace{+ \cdots +}_{(N + M_z) / 2 - 1} - + \underbrace{- \cdots -}_{(N - M_z) / 2 - 1} 0 \cdots 0}$, which is orthogonal to $\ket{g_1}$, and consider the corresponding $\mathcal{H}_2$.
Continuing this process, we can construct $\mathcal{H}_1$, $\mathcal{H}_2$, $\cdots$, $\mathcal{H}_\mathcal{M}$ that are orthogonal to each other and satisfy $\mathcal{H} = \bigoplus_{m = 1}^{\mathcal{M}} \mathcal{H}_m$, where $\mathcal{H}$ is the partial Fock space specified by $(L, N, M_z)$, and $\mathcal{M} \in \mathbb{N}$ is determined by $(L, N, M_z)$.
Finally, we assign $S = m$ to the obtained $\mathcal{H}_m$ ($m = 1, 2, \cdots, \mathcal{M}$).
For example, for $(L, N, M_z) = (7, 6, 0)$, we obtain $\mathcal{M} = 4$, $\ket{g_1} = \ket{+ + + - - - 0}$, $\ket{g_2} = \ket{+ + - + - - 0}$, $\ket{g_3} = \ket{+ + - - + - 0}$, and $\ket{g_4} = \ket{+ - + - + - 0}$, and can construct $\{ \mathcal{H}_m \}_{m = 1}^4$, accordingly.

In a partial Fock space specified by the three quantum numbers, $(N, M_z, S)$, the Perron-Frobenius theorem holds for the matrix representation of $\hat{H}$ using the Fock bases.
Thus, we first separately consider the ground state in each subspace specified by $(N, M_z, S)$ and then examine the state where the ground-state energy is minimal with respect to $M_z$ and $S$ for a fixed $N$.

In the partial Fock space specified by $(N, M_z, S)$, we divide $\hat{H}$ as $\hat{H} = \hat{H}_0 + \hat{H}_1$, where $\hat{H}_0$ is the Hermitian part defined as
\begin{equation}
    \hat{H}_0 := \hat{H}_\mathrm{hop} = - t \sum_{i = 1}^L \sum_{s = \pm} (\hat{a}_{i + 1, s}^\dag \hat{a}_{i, s} + \hat{a}_{i, s}^\dag \hat{a}_{i + 1, s}),
    \label{Eq:model_unperturbed}
\end{equation}
and $\hat{H}_1$ is regarded as the anti-Hermitian part defined as
\begin{equation}
    \hat{H}_1 := \hat{H}_\mathrm{act} = - \varepsilon t \sum_{i = 1}^L \sum_{s = \pm} s (\hat{a}_{i + 1, s}^\dag \hat{a}_{i, s} - \hat{a}_{i, s}^\dag \hat{a}_{i + 1, s}).
    \label{Eq:model_perturbation}
\end{equation}
We denote the ground-state eigenvalue and eigenvector of $\hat{H}_0$ as $E_\mathrm{GS}^{(0)}$ and $\ket{\psi_\mathrm{GS}^{(0)}}$, and the ground-state eigenvalue and eigenvector of $\hat{H}$ as $E_\mathrm{GS}$ and $\ket{\psi_\mathrm{GS}}$, respectively.

From the generic property of the eigenvalues of non-Hermitian matrices combined with the fact that $\hat{H}$ and $\hat{H}_0$ satisfy the Perron-Frobenius condition, we can show that
\begin{equation}
    E_\mathrm{GS} \geq E_\mathrm{GS}^{(0)},
\end{equation}
where the equality is satisfied if and only if $\hat{H}_1 \ket{\psi_\mathrm{GS}} = 0$ (see Appendix~\ref{App:bound}).
Thus, the ground-state energy generically increases as we introduce nonzero $\varepsilon$, except for the special case with $\hat{H}_1 \ket{\psi_\mathrm{GS}} = 0$.
We can also show that $E_\mathrm{GS}^{(0)}$ does not depend on the quantum numbers $M_z$ and $S$ (see Appendix~\ref{App:gs_property_nonfull}).
In the following, we show that when the activity is finite, $\varepsilon > 0$, we have $E_\mathrm{GS} = E_\mathrm{GS}^{(0)}$ for the fully ferromagnetic ground states (i.e., with $M_z = \pm N$) and $ E_\mathrm{GS} > E_\mathrm{GS}^{(0)}$ otherwise, which suggests that ferromagnetism is stabilized by activity for $h_x = J_z = 0$.

\subsection{Fully ferromagnetic states are robust against activity}
\label{Subsec:proof_ferro}

For the fully ferromagnetic states, we focus on the case with $M_z = N$ without loss of generality.
In this case, $S = 1$ since the spin configuration of the particles except for the empty sites is unique.

Within the partial Fock space specified by $(N, M_z = N , S = 1)$, $\hat{H}_0$ and $\hat{H}_1$ are reduced to $\hat{H}_0 = -t \sum_{i = 1}^L (\hat{a}_{i + 1, +}^\dag \hat{a}_{i, +} + \hat{a}_{i, +}^\dag \hat{a}_{i + 1, +})$ and $\hat{H}_1 = -\varepsilon t \sum_{i = 1}^L (\hat{a}_{i + 1, +}^\dag \hat{a}_{i, +} - \hat{a}_{i, +}^\dag \hat{a}_{i + 1, +})$, respectively.
Since $\hat{H}_0$ and $\hat{H}_1$ are commutable, $\ket{\psi_\mathrm{GS}^{(0)}}$ is also an eigenvector of $\hat{H}_1$.
We denote the corresponding eigenvalue of $\hat{H}_1$ as $E_1$, which is purely imaginary since $\hat{H}_1$ is anti-Hermitian.
Thus, we obtain $\hat{H} \ket{\psi_\mathrm{GS}^{(0)}} = (\hat{H}_0 + \hat{H}_1) \ket{\psi_\mathrm{GS}^{(0)}} = (E_\mathrm{GS}^{(0)} + E_1) \ket{\psi_\mathrm{GS}^{(0)}}$, which means that $\ket{\psi_\mathrm{GS}^{(0)}}$ and $E_\mathrm{GS}^{(0)} + E_1$ are the eigenvector and eigenvalue of $\hat{H}$, respectively.
Since $\ket{\psi_\mathrm{GS}^{(0)}}$ is the Perron-Frobenius eigenvector for $\hat{H}_0$, $\ket{\psi_\mathrm{GS}^{(0)}}$ is also the Perron-Frobenius eigenvector for $\hat{H}$ (i.e., $\ket{\psi_\mathrm{GS}^{(0)}} = \ket{\psi_\mathrm{GS}}$), and the corresponding eigenvalue $E_\mathrm{GS}=E_\mathrm{GS}^{(0)} + E_1$ must be real. This means that $E_\mathrm{GS}=E_\mathrm{GS}^{(0)}$, and correspondingly,  $\hat{H}_1 \ket{\psi_\mathrm{GS}} = 0$.

In Appendix~\ref{App:gs_property_full}, we show another proof of $\hat{H}_1 \ket{\psi_\mathrm{GS}} = 0$ using the Jordan-Wigner transformation~\cite{Giamarchi2004}.

\subsection{Activity-induced energy increase in other ground states}

Assuming $2 \leq N \leq L - 1$ and $|M_z| \leq N - 2$, we consider a partial Fock space specified by $(N, M_z, S)$.
Given $|M_z| \leq N - 2$, we should be able to find $\ket{f_1} := \ket{+ s_1 s_2 \cdots s_{N - 2} \, 0 - \underbrace{0 \cdots 0}_{L - N - 1}}$ and $\ket{f_2} := \ket{0 \, s_1 s_2 \cdots s_{N - 2} - \underbrace{0 \cdots 0}_{L - N - 1} +}$ in this space, where $\{ s_n \}_{n = 1}^{N - 2}$ $(s_n \in \{ +, - \})$ represent a sequence of spins that does not include 0.

Now, we can check that for $\ket{f_0} := \ket{+ s_1 s_2 \cdots s_{N - 2} - \underbrace{0 \cdots 0}_{L - N} }$, only $\braket{f_0 | \hat{H}_1 | f_1}$ and $\braket{f_0 | \hat{H}_1 | f_2}$ are non-zero ($= - \varepsilon$), and for any state $\ket{\phi}$ that is orthogonal to both $\ket{f_1}$ and $\ket{f_2}$, it should be $\braket{f_0 | \hat{H}_1 | \phi}=0$.
According to the Perron-Frobenius theorem, we can expand $\ket{\psi_\mathrm{GS}}$ as $\ket{\psi_\mathrm{GS}} = c_1 \ket{f_1} + c_2 \ket{f_2} + \ket{\phi}$ with $c_1, c_2 > 0$.
Thus, for $\varepsilon > 0$, we obtain $\braket{f_0 | \hat{H}_1 | \psi_\mathrm{GS}} = -\varepsilon (c_1 + c_2) \neq 0$, which means $\hat{H}_1 \ket{\psi_\mathrm{GS}} \neq 0$, and therefore $E_\mathrm{GS} >  E_\mathrm{GS}^{(0)}$ (see Appendix~\ref{App:bound}).

\section{Two-particle problem}
\label{Sec:two_particle_problem}

Applying the results of Sec.~\ref{Sec:proof} to two-particle systems ($N = 2$ and $L \geq 3$), we can show that a finite $\varepsilon$ stabilizes the ferromagnetic ground states with $M_z = \pm 2$ compared to the paramagnetic ground state with $M_z = 0$.
This suggests that the essential effect of activity on the ground state can be captured by the two-particle problem.
In this section, we assume $h_x = J_z = 0$ as in Sec.~\ref{Sec:proof} and obtain the ground-state properties of two-particle systems explicitly to gain further insight.

\subsection{Ferromagnetic ground states}
\label{Subsec:two_particle_ferro}

For the ferromagnetic states, we focus on the case with $M_z = 2$ without loss of generality.
We consider a unitary transformation $\hat{T}$ that represents the one-site translation of all the particles; for example, $\hat{T} \ket{0 + 0 \, 0 \, +} = \ket{+ \, 0 + 0 \, 0}$.
Since $\hat{T}$ commutes with $\hat{H}$, the ground-state eigenvector $\ket{\psi_\mathrm{GS}}$ is an eigenvector of $\hat{T}$, i.e., $\hat{T} \ket{\psi_\mathrm{GS}} = T_\mathrm{GS} \ket{\psi_\mathrm{GS}}$ with $|T_\mathrm{GS}| = 1$.
On the other hand, expanding $\ket{\psi_\mathrm{GS}}$ as $\ket{\psi_\mathrm{GS}} = \sum_n c_n \ket{f_n}$, where $\{ \ket{f_n} \}_n$ is the set of Fock bases, we can take $c_n > 0$ for any $n$, according to the Perron-Frobenius theorem.
Then, $\hat{T} \ket{\psi_\mathrm{GS}} = \sum_n c'_n \ket{f_n}$ with $c'_n > 0$ for any $n$, and thus we obtain $T_\mathrm{GS} = 1$, which means that the ground state is translationally invariant.

Considering the translational invariance of the ground state, we can expand $\ket{\psi_\mathrm{GS}}$ as $\ket{\psi_\mathrm{GS}} = \sum_{l} d_l \ket{l}$, where $\{ \ket{l} \}_l$ is the set of translationally invariant bases defined as
\begin{equation}
    \left\{
    \begin{array}{l}
        \ket{1} := L^{-1 / 2} (\ket{+ + 0 \cdots 0} + \text{t.c.}) \\
        \ket{2} := L^{-1 / 2} (\ket{+ \, 0 + 0 \cdots 0} + \text{t.c.}) \\
        \ \ \ \ \ \ \ \vdots \\
        \ket{(L - 1) / 2} := L^{-1 / 2} (\ket{+ \underbrace{0 \cdots 0}_{(L - 3) / 2} + \, 0 \cdots 0} + \text{t.c.})
    \end{array}
    \right.
\end{equation}
for odd $L$ and 
\begin{equation}
    \left\{
    \begin{array}{l}
        \ket{1} := L^{-1 / 2} (\ket{+ + 0 \cdots 0} + \text{t.c.}) \\
        \ket{2} := L^{-1 / 2} (\ket{+ \, 0 + 0 \cdots 0} + \text{t.c.}) \\
        \ \ \ \ \ \ \ \vdots \\
        \ket{L / 2 - 1} := L^{-1 / 2} (\ket{+ \underbrace{0 \cdots 0}_{L / 2 - 2} + \, 0 \cdots 0} + \text{t.c.}) \\
        \ket{L / 2} := (L / 2)^{-1 / 2} (\ket{+ \underbrace{0 \cdots 0}_{L / 2 - 1} + \, 0 \cdots 0} + \text{t.c.})
    \end{array}
    \right.
\end{equation}
for even $L$.
Here, $\text{t.c.}$ is the summation of all the Fock bases that are connected to the term just before $\text{t.c.}$ by translation, i.e., single or multiple operation(s) of $\hat{T}$; for example, for $L = 4$, $\ket{1} = 2^{-1} (\ket{+ + 0 \, 0} + \ket{0 + + \, 0} + \ket{0 \, 0 + +} + \ket{+ \, 0 \, 0 \, +})$ and $\ket{2} = 2^{-1 / 2} (\ket{+ \, 0 + 0} + \ket{0 + 0 \, +})$.
Since $\sum_{i = 1}^L \sum_{s = \pm} (\hat{a}_{i + 1, s}^\dag \hat{a}_{i, s} - \hat{a}_{i, s}^\dag \hat{a}_{i + 1, s}) \ket{l} = 0$ for any $l$, which means that the effect of $\varepsilon$ cancels out for any $\ket{l}$, we obtain $\hat{H}_1 \ket{\psi_\mathrm{GS}} = 0$, consistent with the general result in Sec.~\ref{Subsec:proof_ferro}.
Note that $\hat{H}_1 \ket{\psi_\mathrm{GS}} = 0$ also holds for the ferromagnetic two-particle ground state in the case with $J_z > 0$ and $h_x = 0$, according to the same type of discussion based on the translational invariance.

Representing $\hat{H}$ with bases $\{ \ket{l} \}_l$, we obtain an $(L - 1) / 2 \times (L - 1) / 2$ matrix,
\begin{equation}
    \hat{H} = -2 t
    \begin{pmatrix}
        0 & 1\\
        1 & 0 & 1 \\
         & 1 & 0 \\
        & & & \ddots \\
        & & & & 0 & 1 \\
        & & & & 1 & 0 & 1 \\
        & & & & & 1 & 1
    \end{pmatrix},
    \label{Eq:two_particle_hamiltonian_ferro_odd}
\end{equation}
for odd $L$ and an $L / 2 \times L / 2$ matrix,
\begin{equation}
    \hat{H} = -2 t
    \begin{pmatrix}
        0 & 1\\
        1 & 0 & 1 \\
         & 1 & 0 \\
        & & & \ddots \\
        & & & & 0 & 1 \\
        & & & & 1 & 0 & \sqrt{2} \\
        & & & & & \sqrt{2} & 0
    \end{pmatrix},
    \label{Eq:two_particle_hamiltonian_ferro_even}
\end{equation}
for even $L$.
Hamiltonian~\eqref{Eq:two_particle_hamiltonian_ferro_odd} or \eqref{Eq:two_particle_hamiltonian_ferro_even} describes the relative motion of two particles with the same spin state in the translationally invariant subspace.
From Eqs.~\eqref{Eq:two_particle_hamiltonian_ferro_odd} and \eqref{Eq:two_particle_hamiltonian_ferro_even}, we can obtain
\begin{equation}
    \ket{\psi_\mathrm{GS}} \propto \sum_{l = 1}^{(L - 3) / 2} \sin (\pi l / L) \ket{l} + \cos [\pi / (2 L)] \ket{(L - 1) / 2}
    \label{Eq:two_particle_gsvector_ferro_odd}
\end{equation}
for odd $L$ and
\begin{equation}
    \ket{\psi_\mathrm{GS}} \propto \sum_{l = 1}^{L / 2 - 1} \sin (\pi l / L) \ket{l} + 2^{-1/2} \ket{L / 2}
    \label{Eq:two_particle_gsvector_ferro_even}
\end{equation}
for even $L$, and
\begin{equation}
    E_\mathrm{GS} = -4 t \cos (\pi / L)
    \label{Eq:two_particle_gsenergy_ferro}
\end{equation}
for both odd $L$ and even $L$.
Note that Eqs.~\eqref{Eq:two_particle_gsvector_ferro_odd}-\eqref{Eq:two_particle_gsenergy_ferro} are also applied to the case with $M_z = - 2$ by replacing spin $+$ with spin $-$.

\subsection{Paramagnetic ground state}
\label{Subsec:two_particle_nonmag}

As in the ferromagnetic states explained in Sec.~\ref{Subsec:two_particle_ferro}, the ground state is translationally invariant for the paramagnetic state.
Thus, we can expand the ground-state eigenvector as $\ket{\psi_\mathrm{GS}} = \sum_{m = 1}^{L - 1} e_{m} \ket{m}$, where $\{ \ket{m} \}_{m = 1}^{L - 1}$ is the set of translationally invariant bases defined as
\begin{equation}
    \left\{
    \begin{array}{l}
        \ket{1} := L^{-1 / 2} (\ket{+ - 0 \cdots 0} + \text{t.c.}) \\
        \ket{2} := L^{-1 / 2} (\ket{+ \, 0 - 0 \cdots 0} + \text{t.c.}) \\
        \ \ \ \ \ \ \ \vdots \\
        \ket{L - 1} := L^{-1 / 2} (\ket{+ \, 0 \cdots 0 \, -} + \text{t.c.}).
    \end{array}
    \right.
    \label{Eq:trans_inv_bases_nonmag}
\end{equation}

We can represent $\hat{H}$ with bases $\{ \ket{m} \}_{m = 1}^{L - 1}$ as
\begin{equation}
    \hat{H} = -2 t
    \begin{pmatrix}
        0 & 1 + \varepsilon\\
        1 - \varepsilon & 0 & 1 + \varepsilon \\
         & 1 - \varepsilon & 0 \\
        & & & \ddots \\
        & & & & 0 & 1 + \varepsilon \\
        & & & & 1 - \varepsilon & 0
    \end{pmatrix},
    \label{Eq:two_particle_hamiltonian_nonmag}
\end{equation}
which describes the relative motion of two particles with opposite spin states in the translationally invariant subspace.
In contrast to the ferromagnetic case [Eqs.~\eqref{Eq:two_particle_hamiltonian_ferro_odd} and \eqref{Eq:two_particle_hamiltonian_ferro_even}], the effect of activity $\varepsilon$ appears in Eq.~\eqref{Eq:two_particle_hamiltonian_nonmag}.
More specifically, reflecting the asymmetric hopping and the prohibited exchange of the two particles, the Hamiltonian~\eqref{Eq:two_particle_hamiltonian_nonmag} is equivalent to the Hatano-Nelson model under the open boundary condition~\cite{Hatano1996}.
Following the standard procedure~\cite{Hatano1996,Yao2018}, using $\hat{V} := \mathrm{diag} (1, [(1 - \varepsilon ) / (1 + \varepsilon)]^{1 / 2}, (1 - \varepsilon ) / (1 + \varepsilon), \cdots, [(1 - \varepsilon ) / (1 + \varepsilon)]^{(L - 2) / 2})$, we can transform Eq.~\eqref{Eq:two_particle_hamiltonian_nonmag} into a Hermitian matrix:
\begin{equation}
    \hat{V}^{-1} \hat{H} \hat{V} = -2 t \sqrt{1 - \varepsilon^2}
    \begin{pmatrix}
        0 & 1 \\
        1 & 0 & 1 \\
         & 1 & 0 \\
        & & & \ddots \\
        & & & & 0 & 1 \\
        & & & & 1 & 0
    \end{pmatrix}.
\end{equation}
Diagonalizing $\hat{V}^{-1} \hat{H} \hat{V}$, we finally obtain
\begin{equation}
    \ket{\psi_\mathrm{GS}} \propto \sum_{m = 1}^{L - 1} [(1 - \varepsilon) / (1 + \varepsilon)]^{m / 2} \sin (\pi m / L) \ket{m}
    \label{Eq:two_particle_gsvector_nonmag}
\end{equation}
and
\begin{equation}
    E_\mathrm{GS} = - 4 t \sqrt{1 - \varepsilon^2} \cos (\pi / L).
    \label{Eq:two_particle_gsenergy_nonmag}
\end{equation}
Comparing Eqs.~\eqref{Eq:two_particle_gsenergy_ferro} and \eqref{Eq:two_particle_gsenergy_nonmag}, we find that the ground-state energy for the ferromagnetic states is lower than that for the paramagnetic state as long as $\varepsilon > 0$.

\begin{figure}[t]
    \centering
    \includegraphics[scale=1]{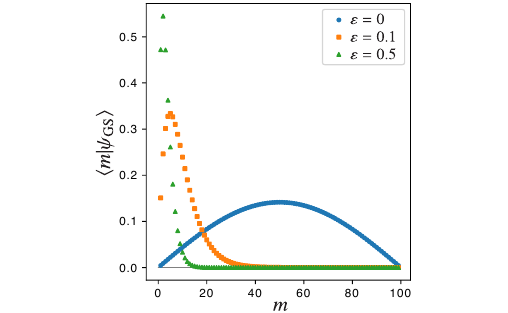}
    \caption{Normalized expansion coefficients $\braket{m | \psi_\mathrm{GS}}$ of the two-particle paramagnetic ground state for $h_x = J_z = 0$ and $L = 100$.
    Index $m$ represents the coordinate of the particle with spin $-$ relative to the other particle with spin $+$.
    In contrast to the case with $\varepsilon = 0$ (blue circles), $\braket{m | \psi_\mathrm{GS}}$ decays exponentially as a function of $m$ for $\varepsilon > 0$ (orange squares and green triangles), reflecting the two-particle bound state.}
    \label{Fig:twoparticle}
\end{figure}

In Fig.~\ref{Fig:twoparticle}, we plot the expansion coefficients in Eq.~\eqref{Eq:two_particle_gsvector_nonmag} with varying $\varepsilon$, which shows exponential decay that reflects the bound state formation for $\varepsilon > 0$.
This indicates the appearance of the so-called non-Hermitian skin effect~\cite{Yao2018} at the two-particle level~\cite{Lee2021}: the particle with the $-$ spin localizes on the right of the particle with the $+$ spin, leading to a two-particle bound state.

\section{Mean-field theory}
\label{Sec:mean_field_theory}

According to the result of the two-particle problem for $h_x = J_z = 0$ (see Sec.~\ref{Sec:two_particle_problem}), the many-body ground state for $h_x > 0$ and/or $J_z > 0$ is expected to reflect the $\varepsilon$-induced bound state of two particles with different spin states.
Indeed, a single-site mean-field theory, which ignores such bound state formation, does not reproduce any $\varepsilon$-dependence of the ground-state energy or eigenvector, as shown in Appendix~\ref{Appsub:single_site_mft}.
Thus, we expect that the bound state is the key to the $\varepsilon$-induced ferromagnetism, which is observed in numerical experiments (see Sec.~\ref{Sec:numerical_results}).
In the following, we consider a two-site mean-field theory to incorporate minimal effects of the bound state and discuss the ferromagnetic transition for $h_x > 0$ and $J_z \geq 0$.
We assume even $L$ in this section.

Our approach to constructing the mean-field theory is to use the variational formulation.
Though the variational formulation is often used in Hermitian systems~\cite{Kleinert2009}, the variational principle is not generically applicable to non-Hermitian systems such as our model of interest, Eq.~\eqref{Eq:model_HCB1lane}.
Nevertheless, in the same spirit as the principle of minimal sensitivity~\cite{Stevenson1981}, we extrapolate the variational principle in the Hermitian regime ($\varepsilon = 0$) to the non-Hermitian regime ($\varepsilon > 0$).
Following the general formulation explained in Appendix~\ref{App:mft}, we write a set of variational parameters as $\alpha$, the variational ground-state eigenvector as $\ket{\tilde{\psi}_\mathrm{GS} (\alpha)}$, and the variational ground-state energy as $\tilde{E}_\mathrm{GS} (\alpha) := \bra{\tilde{\phi}_\mathrm{GS} (\alpha)} \hat{H} \ket{\tilde{\psi}_\mathrm{GS} (\alpha)}$, where $\ket{\tilde{\phi}_\mathrm{GS} (\alpha)}$ is the left eigenvector corresponding to $\ket{\tilde{\psi}_\mathrm{GS} (\alpha)}$.
Using the optimized variational parameters, $\alpha^* := \arg \min_\alpha \tilde{E}_\mathrm{GS} (\alpha)$, we can obtain the ground-state energy and eigenvector at the mean-field level as $E_\mathrm{GS}^\mathrm{MF} := \tilde{E}_\mathrm{GS} (\alpha^*)$ and $\ket{\psi_\mathrm{GS}^\mathrm{MF}} := \ket{\tilde{\psi}_\mathrm{GS} (\alpha^*)}$, respectively.

To determine the specific form of $\ket{\tilde{\psi}_\mathrm{GS} (\alpha)}$, we consider a two-site mean-field theory.
We divide the whole system into $L / 2$ clusters of two consecutive sites.
To each cluster specified by odd $i \in \{ 1, 3, \cdots, L - 1 \}$, we assign a mean-field Hamiltonian:
\begin{align}
    \hat{\tilde{H}}_{i} (\alpha) := & -t \sum_{s = \pm} (\hat{a}_{i + 1, s}^\dag \hat{a}_{i, s} + \hat{a}_{i, s}^\dag \hat{a}_{i + 1, s}) \nonumber \\
    & - \varepsilon t \sum_{s = \pm} s (\hat{a}_{i + 1, s}^\dag \hat{a}_{i, s} - \hat{a}_{i, s}^\dag \hat{a}_{i + 1, s}) \nonumber \\
    & - h_x \sum_{s = \pm} (\hat{a}_{i, -s}^\dag \hat{a}_{i, s} + \hat{a}_{i + 1, -s}^\dag \hat{a}_{i + 1, s}) - J_z \hat{m}_i^z \hat{m}_{i + 1}^z \nonumber \\
    & - \mu (\hat{n}_i + \hat{n}_{i + 1}) - \nu (\hat{m}_i^z + \hat{m}_{i + 1}^z) \nonumber \\
    & - \sum_{s = \pm} \lambda_s (\hat{a}_{i, s} + \hat{a}_{i, s}^\dag + \hat{a}_{i + 1, s} + \hat{a}_{i + 1, s}^\dag),
\end{align}
where $\hat{n}_i := \hat{n}_{i, +} + \hat{n}_{i, -}$.
We also define the total mean-field Hamiltonian as $\hat{\tilde{H}} (\alpha) := \sum_{\text{odd } i} \hat{\tilde{H}}_i (\alpha)$.
Here, $\alpha := \{ \nu, \lambda_+, \lambda_- \}$ is a set of variational parameters: ferromagnetic mean field $\nu$ and superfluid mean field $\lambda_s$ for particles with spin $s$.
We assume $\lambda_s \geq 0$ so that the Perron-Frobenius theorem holds for $\hat{\tilde{H}}_{i} (\alpha)$ as well as Hamiltonian~\eqref{Eq:model_HCB1lane}.
Chemical potential $\mu$ is also introduced to keep the total particle number to $N$ on average.

Writing the ground-state eigenvector and left eigenvector of $\tilde{H}_i (\alpha)$ as $\ket{\tilde{\psi}_{\mathrm{GS}, i} (\alpha)}$ and $\ket{\tilde{\phi}_{\mathrm{GS}, i} (\alpha)}$, respectively, we obtain the ground-state eigenvectors for $\hat{\tilde{H}} (\alpha)$ as the tensor products:
\begin{equation}
    \left\{
    \begin{array}{l}
        \displaystyle \ket{\tilde{\psi}_\mathrm{GS} (\alpha)} = \bigotimes_{\text{odd } i} \ket{\tilde{\psi}_{\mathrm{GS}, i} (\alpha)} \vspace{1mm} \\
        \displaystyle \ket{\tilde{\phi}_\mathrm{GS} (\alpha)} = \bigotimes_{\text{odd } i} \ket{\tilde{\phi}_{\mathrm{GS}, i} (\alpha)}.
    \end{array}
    \right.
\end{equation}
Then, following the method explained above, we obtain the variational ground-state energy as $\tilde{E}_\mathrm{GS} (\alpha) = \braket{\tilde{\phi}_\mathrm{GS} (\alpha) | \hat{H} | \tilde{\psi}_\mathrm{GS} (\alpha)}$, which is reduced to
\begin{align}
    \frac{2 \tilde{E}_\mathrm{GS} (\alpha)}{L} = & - \sum_{s = \pm} t [(1 + \varepsilon s) \braket{\hat{a}_{1, s}^\dag} \braket{\hat{a}_{2, s}} + (1 - \varepsilon s) \braket{\hat{a}_{2, s}^\dag} \braket{\hat{a}_{1, s}}] \nonumber \\
    & - J \braket{\hat{m}_1^z} \braket{\hat{m}_2^z} + \nu (\braket{\hat{m}_1^z} + \braket{\hat{m}_2^z}) \nonumber \\
    & + \sum_{s = \pm} \lambda_s (\braket{\hat{a}_{1, s}} + \braket{\hat{a}_{1, s}^\dag} + \braket{\hat{a}_{2, s}} + \braket{\hat{a}_{2, s}^\dag}) + \tilde{\varepsilon}_\mathrm{GS},
\end{align}
where $\tilde{\varepsilon}_\mathrm{GS}$ is the ground-state energy of the two-site Hamiltonian $\hat{\tilde{H}}_i (\alpha)$, and $\braket{\cdots} := \braket{\tilde{\phi}_\mathrm{GS} (\alpha) | \cdots | \tilde{\psi}_\mathrm{GS} (\alpha)}$.
To obtain the ground-state eigenvector at the mean-field level from $\ket{\psi_\mathrm{GS}^\mathrm{MF}} = \ket{\tilde{\psi}_\mathrm{GS} (\alpha^*)}$ with $\alpha^* = \arg \min_\alpha \tilde{E}_\mathrm{GS} (\alpha)$, we minimize $\tilde{E}_\mathrm{GS} (\alpha)$ using scipy.optimize.dual\_annealing in the SciPy package~\cite{Scipy2020}.

\begin{figure}[t]
    \centering
    \includegraphics[scale=1]{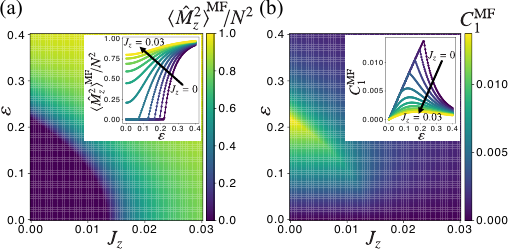}
    \caption{Magnetic phase diagram obtained by the two-site mean-field theory for $t = 1$, $h_x = 0.01$, and $\rho = 0.5$.
    (a) Heatmap of the normalized squared total magnetization $\braket{{\hat{M}_z}^2}^\mathrm{MF} / N^2$ in the $J_z$-$\varepsilon$ plane.
    The inset shows the $\varepsilon$-dependence of $\braket{{\hat{M}_z}^2}^\mathrm{MF} / N^2$, which indicates the ferromagnetic transition, for several values of $J_z$ from $0$ to $0.03$.
    (b) Heatmap of the mean-field binding strength $C_1^\mathrm{MF}$.
    The inset shows the $\varepsilon$-dependence of $C_1^\mathrm{MF}$, which shows a peak at the $\varepsilon$-induced transition point.}
    \label{Fig:meanfield}
\end{figure}

In Fig.~\ref{Fig:meanfield}(a), we plot the heatmap of the obtained normalized squared total magnetization, $\braket{{\hat{M}_z}^2}^\mathrm{MF} / N^2 := \braket{\psi_\mathrm{GS}^\mathrm{MF} | {\hat{M}_z}^2 | \psi_\mathrm{GS}^\mathrm{MF}} / N^2$, for the parameter sets ($t = 1$, $h_x = 0.01$, and $\rho = 0.5$) used in the numerical study (Sec.~\ref{Sec:numerical_results}).
This figure shows that the system undergoes the ferromagnetic transition when $\varepsilon$ or $J_z$ is increased, as observed in the numerical study [see Fig.~\ref{Fig:PD}(a)].
Furthermore, the mean-field counterpart of the binding strength, $C_1^\mathrm{MF} := L^{-1} \sum_{i = 1}^L \braket{\psi_\mathrm{GS}^\mathrm{MF} | \hat{n}_{i, +} \hat{n}_{i+1, -} - \hat{n}_{i, -} \hat{n}_{i+1, +} | \psi_\mathrm{GS}^\mathrm{MF}}$, shows a peak at the transition point as a function of $\varepsilon$ [Fig.~\ref{Fig:meanfield}(b)], as seen in the numerical study [Fig.~\ref{Fig:PD}(b)].
Thus, the two-site mean-field theory qualitatively reproduces the magnetic properties of the ground state for model~\eqref{Eq:model_HCB1lane}.
Our results confirm that $\ket{\psi_\mathrm{GS}^\mathrm{MF}}$ reflects the $\varepsilon$-induced two-particle bound state and corroborate the conjecture that the bound state is essential to the $\varepsilon$-induced ferromagnetism.

Focusing on $J_z = 0$, we can explain the $\varepsilon$-induced ferromagnetic transition and $\varepsilon$-dependence of $C_1^\mathrm{MF}$ as a consequence of the competition between the transverse magnetic field $h_x$, which favors the paramagnetic phase, and the activity $\varepsilon$, which favors the ferromagnetic phase as we discussed in Sec.~\ref{Sec:proof}.
In the paramagnetic phase, as suggested by the paramagnetic state of two particles (see Fig.~\ref{Fig:twoparticle}), the binding strength $C_1^\mathrm{MF}$ and ground-state energy are expected to increase as $\varepsilon$ increases.
The ferromagnetic transition will occur when the energy in the paramagnetic phase exceeds the energy in the ferromagnetic phase.
In the ferromagnetic phase, particle pairs with different spin states (i.e., $+$ and $-$) are less likely to appear as the ferromagnetic order is enhanced by $\varepsilon$, leading to the decrease of $C_1^\mathrm{MF}$.

We note a few remarks on the properties of our mean-field theory.
First, the inset of Fig~\ref{Fig:meanfield}(a) suggests that $\braket{{\hat{M}_z}^2}^\mathrm{MF}$ linearly increases as a function of $\varepsilon - \varepsilon_c^\mathrm{MF} (J_z)$ near $\varepsilon = \varepsilon_c^\mathrm{MF} (J_z)$, where $\varepsilon_c^\mathrm{MF} (J_z)$ is the $J_z$-dependent mean-field critical point.
Thus, the mean-field critical exponent for magnetization at the $\varepsilon$-induced ferromagnetic transition is $\beta^\mathrm{MF} = 1/2$ as obtained by the standard mean-field theory~\cite{Chaikin1995}.
Further studies are necessary to elucidate whether the transition is indeed continuous for the original model~\eqref{Eq:model_HCB1lane} and how the critical exponents can deviate from the mean-field values.
Second, the mean-field nature of the theory leads to a nonzero superfluid order parameter, which can be zero in the thermodynamic limit of model~\eqref{Eq:model_HCB1lane}, according to the expectation from the quasi-long-range order typical to 1D systems~\cite{Giamarchi2004}.
To improve this point, other methods such as the Bethe ansatz will be required.

\section{Two-lane model}

\begin{figure}
    \centering
    \includegraphics[width=8.5cm]{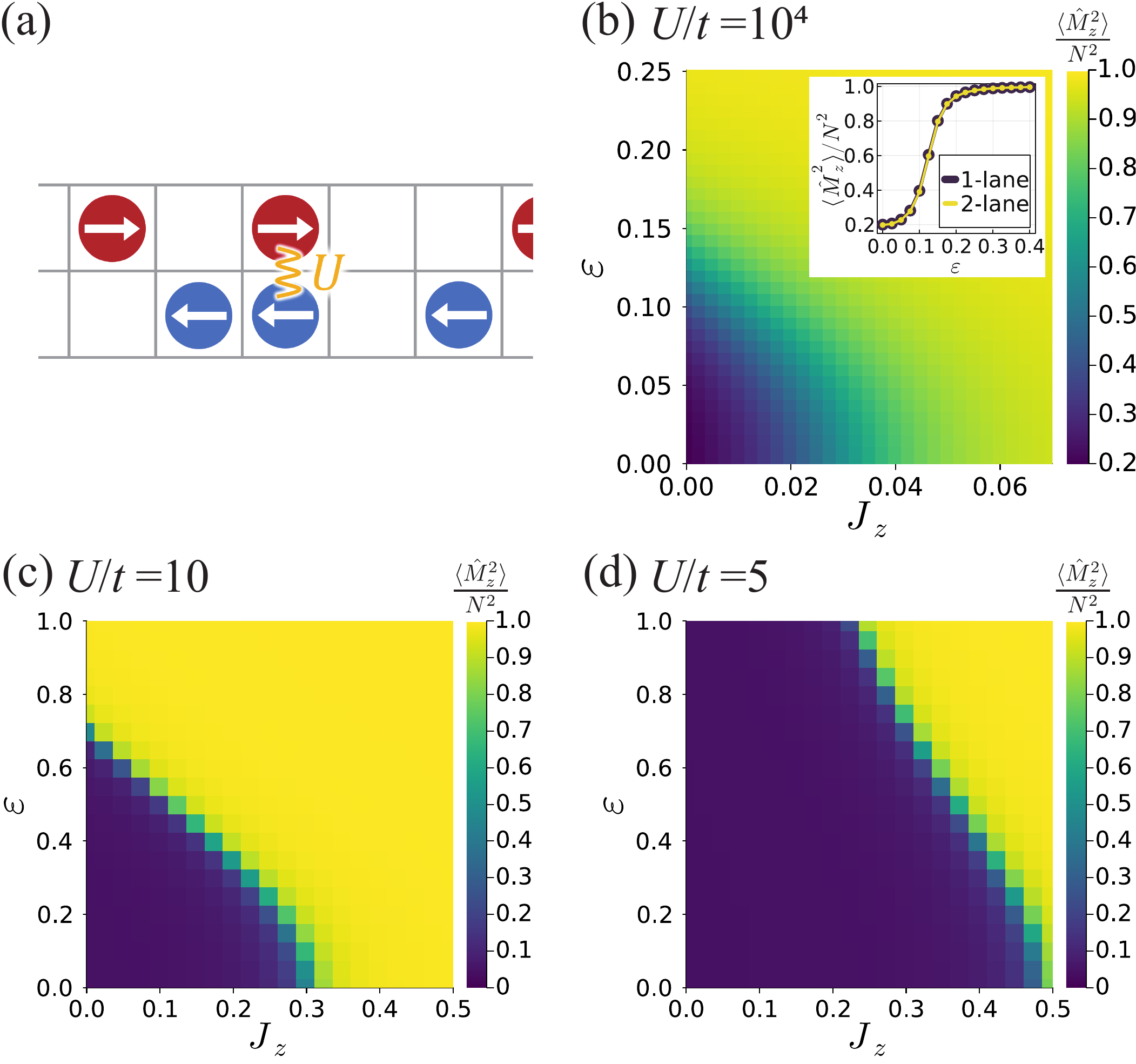}
    \caption{(a) Two-lane model [Eq.~\eqref{Eq:model_HCB2lane}]. $+$ ($-$) particles run only on the upper (lower) lane. There is a repulsive Hubbard-type interaction $\hat{H}_\mathrm{Hub}$ [Eq.~\eqref{eq:H_Hub}] between the lanes. (b-d) Normalized squared magnetization for the two-lane model with (b) $U=10^4$, (c) $U=10$, and (d) $U=5$. The inset of the panel (b) is the normalized squared magnetization at $J_z=0.0$. For comparison, the same quantity for the one-lane model \eqref{Eq:model_HCB1lane} is also plotted. For the panels (b)-(d), we set the parameters as $L=10$, $\rho=0.5$, $t=1.0$, and $h_x=0.01$. }
    \label{fig:2lane}
\end{figure}

In this section, we consider relaxing the strict hard-core condition to see the robustness of activity-induced ferromagnetism. For this purpose, we examine the case where we allow the $+$ and $-$ particles to sit on the same site. This is realized by replacing the projection operator in the Hamiltonian \eqref{Eq:model_HCB1lane} and introducing
\begin{equation}
\hat{H}_\mathrm{2lane} := \hat{P}_{n_+, n_- \leq 1} \Big( \hat{H}_\mathrm{hop} + \hat{H}_\mathrm{act} + \hat{H}_\mathrm{TFIM} + \hat{H}_\mathrm{Hub} \Big) \hat{P}_{n_+, n_- \leq 1}, \label{Eq:model_HCB2lane}  
\end{equation}
where
\begin{equation}
    \hat{H}_\mathrm{Hub}:= U \sum_{i=1}^L \hat{n}_{i,+} \hat{n}_{i,-}. \label{eq:H_Hub}
\end{equation}
Here, $\hat{P}_{n_+, n_- \leq 1}:=\hat{P}_{n \leq 1, +}\hat{P}_{n \leq 1, -}$ and $\hat{P}_{n \leq 1, s}$ is a projection operator to the Hilbert space where the number of particles with $s \in \{ +, - \}$ is not greater than one. This projection prohibits double occupancy for each single component. We call this the two-lane model, since it can be regarded as a set of two ``lanes" where each component runs in a single lane as in Fig~\ref{fig:2lane}(a). For comparison, we call our main model [Eq.~\eqref{Eq:model_HCB1lane}] a one-lane model in this subsection. In the strong coupling limit ($U/t \to \infty$), the two-lane model is reduced to the one-lane model.

The numerical results of the exact diagonalization for the two-lane model are summarized in Figs.~\ref{fig:2lane}(b)-(d). Figure~\ref{fig:2lane}(b) is for very large $U$ ($=10^4 t$) and the results nicely agree with ones for the one-lane model shown in Fig.~\ref{Fig:PD}(a). Figures~\ref{fig:2lane}(c) and (d) are for still large but more realistic values of $U$. The activity-induced ferromagnetic order appears for both cases, which means that the strict hard-core condition is not required to realize the activity-induced ferromagnetism. On the other hand, we need larger $\varepsilon$ and $J_z$ to achieve the phase transition for these cases. In particular, the transition point disappears at $J_z=0$ for $U/t=5$. This is consistent with the analysis in the previous sections, which shows that the strong repulsion induces a two-particle bound state, which plays an important role in stabilizing the ferromagnetic order.

\section{Summary and Outlook}

Here we have demonstrated that in a 1D model of bosons with ferromagnetic interactions, the ferromagnetic (i.e., flocking) phase is enhanced by non-Hermiticity (i.e., activity). We further found that this ferromagnetic phase survives even without the ferromagnetic interactions, which is supported by the proof that activity generically increases the ground state energy of paramagnetic states, as well as the mean-field theory. Although flocking appears typically in active systems with aligning interactions~\cite{Vicsek1995}, there are several examples that show large-scale velocity alignment without explicit aligning interactions~\cite{henkes2020dense,caprini2023flocking,kopp2023spontaneous}. The mechanism of flocking in the quantum model that we found seems distinct from these classical situations; compared with a classical model of MIPS, the repulsive interactions between the bosons are enhanced in the sense of bias in the ensembles~\cite{Adachi2022}, rather than being weakened or turning into attractive.

We have shown that for the observation of activity-induced ferromagnetism, spin-dependent asymmetric hopping [Eq.~\eqref{Eq:H_act}] added to the two-component Bose-Hubbard model is sufficient, and the ferromagnetic interaction, as well as the hard-core condition, is likely unnecessary. Components required for this setup have already been realized in experiments; two-component Bose gas (e.g.,$^{87}$Rb) has been extensively studied~\cite{Gross2017} and the spin-independent asymmetric hopping has been realized using a dissipative optical lattice~\cite{gou2020tunable, liang2022dynamic}. A challenge may lie in detecting the ferromagnetic state. The most direct way is to realize the ground state through the single trajectory dynamics and measure the spin degrees of freedom, which may be difficult for many particle systems. An alternative way is to observe the loss dynamics. A recent theoretical study has revealed that the quantum phase transition of decaying eigenmodes in the Lindbladian spectrum is detectable through the relaxation dynamics~\cite{Haga2023}. We expect that a similar phase transition can also be observed in our setup to observe the activity-induced ferromagnetism.

\textit{Note added---.} 
While we were finalizing the manuscript, we noticed an interesting preprint by R. Khasseh et al. \cite{Khasseh2023} which studies a similar model of 1D quantum active matter with a different approach.

\begin{acknowledgments}
We thank Samuel E. Begg, Shunsuke Furukawa, Ryo Hanai, Masaya Nakagawa, Tomohiro Sasamoto, and Kenji Shimomura for valuable discussions. We especially thank Hosho Katsura for pointing out that the proof of the ferromagnetic ground state (Sec.~\ref{Sec:proof}) can be extended to the non-perturbative regime. The work of K. T. was supported by JSPS KAKENHI Grant No.~JP22K20350 and No.~JP23K17664 and by JST PRESTO Grant No.~JPMJPR2256. The work of K. A. was supported by JSPS KAKENHI Grant No.~JP20K14435. The work of K. K. was supported by JSPS KAKENHI Grant No.~JP19H05795, JP21H01007, and JP23H00095. 
\end{acknowledgments}

\appendix

\section{Bound on the eigenvalues of non-Hermitian matrices}
\label{App:bound}

Let a square matrix $\hat{H}$ be split into its Hermitian part and its anti-Hermitian part, $\hat{H}=\hat{H}_s+\hat{H}_a$.
For an eigenvalue $E$ and eigenvector $\ket{\psi}$ of $\hat{H}$, $\hat{H} \ket{\psi} = (\hat{H}_s+\hat{H}_a) \ket{\psi} = E \ket{\psi}$, we have:
\begin{equation}
    \bra{\psi} \hat{H}_s\ket{\psi} + \bra{\psi}\hat{H}_a \ket{\psi} = E, \label{Eq:RealIm}
\end{equation}
assuming $\braket{\psi | \psi} =1$.
Since $\bra{\psi} \hat{H}_s\ket{\psi}$ is real and $\bra{\psi} \hat{H}_a\ket{\psi}$ is purely imaginary, we find ${\rm Re}(E)=\bra{\psi} \hat{H}_s\ket{\psi}$ and ${\rm Im}(E)=\bra{\psi} \hat{H}_a\ket{\psi}/i $. 
Due to the min-max theorem, we have
\begin{equation}
    E^s_{\rm min} \leq  \bra{\psi} \hat{H}_s\ket{\psi} \leq E^s_{\rm max}, \label{Eq:MinMax}
\end{equation}
where $E^s_{\rm min}$ and $E^s_{\rm max}$ are the smallest and largest eigenvalues of $\hat{H}_s$, respectively. Therefore,
\begin{equation}
    E^s_{\rm min} \leq  {\rm Re}(E) \leq E^s_{\rm max}. \label{Eq:Bendixon}
\end{equation}
This is known as Bendixson's inequality~\cite{householder2013theory,loghin2006bounds}.

If $\hat{H}$ satisfies the Perron-Frobenius condition, then its dominant eigenvalue $E_\mathrm{GS}$ should be real and unique, with an eigenvector $\ket{\psi_\mathrm{GS}}$.
In order to satisfy $E_\mathrm{GS} = E^s_{\rm min}$, we need the first equality in Eq.~\eqref{Eq:MinMax} to hold, which is achieved if and only if $\hat{H}_s\ket{\psi_\mathrm{GS}} = E^s_{\rm min}\ket{\psi_\mathrm{GS}}$, in which case $(\hat{H} - \hat{H}_s)\ket{\psi_\mathrm{GS}} = \hat{H}_a\ket{\psi_\mathrm{GS}} = 0$. Therefore, if $\hat{H}_a\ket{\psi_\mathrm{GS}} \neq 0$, then $E^s_{\rm min} < E_\mathrm{GS}$.
On the other hand, since $\hat{H}_s$ also satisfies the Perron-Frobenius condition, we can say $E^s_{\rm min} = E_\mathrm{GS}$ if $\hat{H}_a \ket{\psi_\mathrm{GS}} = 0$. This is because $\hat{H}_a \ket{\psi_\mathrm{GS}} = 0$ means $\hat{H} \ket{\psi_\mathrm{GS}} = \hat{H}_s \ket{\psi_\mathrm{GS}} = E_\mathrm{GS} \ket{\psi_\mathrm{GS}}$, which suggests that $\ket{\psi_\mathrm{GS}}$ and $E_\mathrm{GS}$ are also the Perron-Frobenius eigenvector and eigenvalue of $\hat{H}_s$, respectively.
Altogether, we have
\begin{equation}
 E_\mathrm{GS}  \geq E^s_{\rm min},
\end{equation}
and
\begin{equation}
\hat{H}_a \ket{\psi_\mathrm{GS}} = 0 \Longleftrightarrow E_\mathrm{GS} = E^s_{\rm min}.
\end{equation}

\section{Ground-state energy for zero activity}
\label{App:gs_property_nonfull}

For a given $N$, we show that the ground-state energy of $\hat{H}_0$ $[= - t \sum_{i = 1}^L \sum_{s = \pm} (\hat{a}_{i + 1, s}^\dag \hat{a}_{i, s} + \hat{a}_{i, s}^\dag \hat{a}_{i + 1, s})]$ [Eq.~\eqref{Eq:model_unperturbed}] does not depend on the quantum numbers $M_z$ and $S$.
For $M_z = N$ (and similarly for $M_z = -N$), $\hat{H}_0$ is reduced to $\hat{H}_0 = -t \sum_{i = 1}^L (\hat{a}_{i + 1, +}^\dag \hat{a}_{i, +} + \hat{a}_{i, +}^\dag \hat{a}_{i + 1, +})$, and thus the matrix representation of $\hat{H}_0$ by the Fock bases is the same as that of the standard hard-core boson model [i.e., $- t \sum_{i = 1}^L (\hat{a}_{i + 1}^\dag \hat{a}_i + \hat{a}_i^\dag \hat{a}_{i + 1})$].

In the following, we assume a partial Fock space specified by $(N, M_z, S)$ with $|M_z| < N$.
We consider a unitary transformation, $\hat{U}$, that represents the translation of all the particles excluding the empty sites; for example, $\hat{U} \ket{+ + - \, 0 \, -} = \ket{- + + \, 0 \, -}$.
Since $\hat{U}$ commutes with $\hat{H}_0$, the ground-state eigenvector $\ket{\psi_\mathrm{GS}^{(0)}}$ is an eigenvector of $\hat{U}$, i.e., $\hat{U} \ket{\psi_\mathrm{GS}^{(0)}} = U_\mathrm{GS}^{(0)} \ket{\psi_\mathrm{GS}^{(0)}}$ with $|U_\mathrm{GS}^{(0)}| = 1$.
On the other hand, expanding $\ket{\psi_\mathrm{GS}^{(0)}}$ as $\ket{\psi_\mathrm{GS}^{(0)}} = \sum_n c_n \ket{f_n}$, where $\{ \ket{f_n} \}_n$ is the set of Fock bases, we can take $c_n > 0$ for any $n$, according to the Perron-Frobenius theorem.
Then, $\hat{U} \ket{\psi_\mathrm{GS}^{(0)}} = \sum_n c'_n \ket{f_n}$ with $c'_n > 0$ for any $n$, and thus we obtain $U_\mathrm{GS}^{(0)} = 1$, which means that $\ket{\psi_\mathrm{GS}^{(0)}}$ is invariant under the translation excluding the empty sites.
We take a set of $\hat{U}$-invariant bases, $\{ \ket{g_m} \}_m$, such that each $\ket{g_m}$ is generated from a single Fock basis by single or multiple operation(s) of $\hat{U}$; for example, a basis generated from $\ket{+ + - \, 0 \, -}$ is $2^{-1} (\ket{+ + - \, 0 \, -} + \ket{- + + \, 0 \, -} + \ket{- - + \, 0 \, +} + \ket{+ - - \, 0 \, +})$.
We can see that the matrix representation of $\hat{H}_0$ by $\{ \ket{g_m} \}_m$ is the same as the matrix representation of the standard hard-core boson model by Fock bases. Related discussions on the energy spectrum of the fermionic Hubbard model are presented in Ref.~\cite{Caspers1989}.

Since the discussion above suggests that the matrix to be diagonalized in obtaining the ground-state energy $E_\mathrm{GS}^{(0)}$ is the same regardless of $M_z$ or $S$, $E_\mathrm{GS}^{(0)}$ does not depend on $M_z$ or $S$.
Specifically, by diagonalizing the standard hard-core boson model~\cite{Giamarchi2004,Rigol2004} (see Appendix~\ref{App:gs_property_full}), we can obtain 
\begin{equation}
E_\mathrm{GS}^{(0)} = - 2 t - 4 t \frac{\cos [(N + 1) \pi / (2 L)] \sin [(N - 1) \pi / (2 L)]}{\sin (\pi / L)}
\end{equation}
for odd $N$ and 
\begin{equation}
E_\mathrm{GS}^{(0)} = - 4 t \frac{\cos [N \pi / (2 L)] \sin [N \pi / (2 L)]}{\sin (\pi / L)}
\end{equation}
for even $N$.

\section{Another proof of robustness against activity in fully ferromagnetic states}
\label{App:gs_property_full}

Assuming $h_x = J_z = 0$, $\varepsilon > 0$, and $2 \leq N \leq L - 1$, we consider the fully ferromagnetic states (i.e., $M_z = \pm N$).
We focus on the case with $M_z = N$ without loss of generality.
Within the partial Fock space specified by $(N, M_z = N , S)$, the two parts of the Hamiltonian, $\hat{H}_0$ and $\hat{H}_1$, are reduced to $\hat{H}_0 = -t \sum_{i = 1}^L (\hat{a}_{i + 1, +}^\dag \hat{a}_{i, +} + \hat{a}_{i, +}^\dag \hat{a}_{i + 1, +})$ and $\hat{H}_1 = -\varepsilon t \sum_{i = 1}^L (\hat{a}_{i + 1, +}^\dag \hat{a}_{i, +} - \hat{a}_{i, +}^\dag \hat{a}_{i + 1, +})$, respectively.
Here, $\hat{H}_0$ is equivalent to the standard hard-core boson model and can be diagonalized by the Jordan-Wigner transformation~\cite{Giamarchi2004,Rigol2004}.
In the following, we rewrite $\hat{a}_{i, +}^{(\dag)}$ as $\hat{a}_i^{(\dag)}$.

Following the standard procedure~\cite{Rigol2004}, we introduce annihilation and creation operators of fermions:
\begin{equation}
    \left\{
    \begin{array}{l}
         \hat{c}_i := \hat{a}_i \prod_{j = 1}^{i - 1} (1 - 2 \hat{a}_j^\dag \hat{a}_j) \vspace{1mm} \\
         \hat{c}_i^\dag := \hat{a}_i^\dag \prod_{j = 1}^{i - 1} (1 - 2 \hat{a}_j^\dag \hat{a}_j).
    \end{array}
    \right.
\end{equation}
Noticing that the total fermion number, $\sum_{i = 1}^L \hat{c}_i^\dag \hat{c}_i$ is equal to the total hard-core boson number, $N = \sum_{i = 1}^L \hat{a}_i^\dag \hat{a}_i$, we can obtain
\begin{equation}
    \hat{H}_0 = -t \sum_{i = 1}^{L - 1} (\hat{c}_{i + 1}^\dag \hat{c}_i + \hat{c}_i^\dag \hat{c}_{i + 1}) + (-1)^N t (\hat{c}_1^\dag \hat{c}_L + \hat{c}_L^\dag \hat{c}_1)
    \label{Eqapp:gs_ferro_h0}
\end{equation}
and
\begin{equation}
    \hat{H}_1 = - \varepsilon t \sum_{i = 1}^{L - 1} (\hat{c}_{i + 1}^\dag \hat{c}_i - \hat{c}_i^\dag \hat{c}_{i + 1}) + (-1)^N \varepsilon t (\hat{c}_1^\dag \hat{c}_L - \hat{c}_L^\dag \hat{c}_1).
    \label{Eqapp:gs_ferro_h1}
\end{equation}

If $N$ is odd, we can diagonalize Eqs.~\eqref{Eqapp:gs_ferro_h0} and \eqref{Eqapp:gs_ferro_h1} with the Fourier transformation, $\hat{c}_j = L^{-1/2} \sum_k \mathrm{e}^{\mathrm{i} k j} \hat{\tilde{c}}_k$ ($k = 2 n \pi / L$ with $n \in \{ 0, 1, \cdots, L - 1 \}$), leading to $\hat{H}_0 + \hat{H}_1 = -2 t \sum_{k} (\cos k - i \varepsilon \sin k) \hat{\tilde{c}}_k^\dag \hat{\tilde{c}}_k$.
Thus, each one-particle state is specified by $k$, and the ground state, $\ket{\psi_\mathrm{GS}}$, is the state in which totally $N$ one-particle states are occupied with the sum of $-2t \cos k$ minimized.
In $\ket{\psi_\mathrm{GS}}$, all the one-particle states are occupied by pairs of $+ |k|$ and $- |k|$ ($\bmod \, L$ for $n$) except for the $k = 0$ state, and thus we obtain $\hat{H}_1 \ket{\psi_\mathrm{GS}} = 2 i \varepsilon t \sum_k \sin k \, \hat{\tilde{c}}_k^\dag \hat{\tilde{c}}_k \ket{\psi_\mathrm{GS}} = 0$.

If $N$ is even, we can diagonalize Eqs.~\eqref{Eqapp:gs_ferro_h0} and \eqref{Eqapp:gs_ferro_h1} with the transformation that respects the anti-periodic boundary condition, $\hat{c}_j = L^{-1/2} \sum_k \mathrm{e}^{\mathrm{i} k j} \hat{\tilde{c}}_k$ [$k = (2 n + 1) \pi / L$ with $n \in \{ 0, 1, \cdots, L - 1 \}$], leading to $\hat{H}_0 + \hat{H}_1 = -2 t \sum_{k} (\cos k - i \varepsilon \sin k) \hat{\tilde{c}}_k^\dag \hat{\tilde{c}}_k$.
Since all the one-particle states are occupied by pairs of $+ |k|$ and $- |k|$ ($\bmod \, L$ for $n$) in $\ket{\psi_\mathrm{GS}}$, we can obtain $\hat{H}_1 \ket{\psi_\mathrm{GS}} = 0$.
Combined with the discussion for odd $N$, we conclude $\hat{H}_1 \ket{\psi_\mathrm{GS}} = 0$ regardless of $N$, which suggests $E_\mathrm{GS} = E^\mathrm{(0)}_\mathrm{GS}$ according to the result of Appendix~\ref{App:bound}.

\section{Non-Hermitian mean-field theory}
\label{App:mft}

We consider a generally non-Hermitian Hamiltonian, $\hat{H}$, that satisfies the conditions for the Perron-Frobenius theorem in a representation with certain bases, such as model~\eqref{Eq:model_HCB1lane}.
We formally introduce the inverse temperature $\beta > 0$.
Defining the thermodynamic potential as $\Omega := -\beta^{-1} \ln \mathrm{Tr} \, e^{-\beta \hat{H}}$, we can obtain the ground-state energy as $E_\mathrm{GS} = \lim_{\beta \to \infty} \Omega$.

We divide $\hat{H}$ into two parts: the main part, $\hat{\tilde{H}} (\alpha)$, where $\alpha$ is a set of variational parameters, and the residual part, $\hat{H} - \hat{\tilde{H}} (\alpha)$.
Expanding $\Omega$ with respect to $\hat{H} - \hat{\tilde{H}} (\alpha)$, we can obtain $E_\mathrm{GS} = \braket{\tilde{\phi}_\mathrm{GS} (\alpha) | \hat{H} | \tilde{\psi}_\mathrm{GS} (\alpha)} + O([\hat{H} - \hat{\tilde{H}} (\alpha)]^2)$ in the limit of $\beta \to \infty$, where $\ket{\tilde{\psi}_\mathrm{GS} (\alpha)}$ and $\ket{\tilde{\phi}_\mathrm{GS} (\alpha)}$ are the ground-state eigenvector and left eigenvector of $\hat{\tilde{H}} (\alpha)$, respectively.
Here, $\ket{\tilde{\phi}_\mathrm{GS} (\alpha)}$ is normalized as $\braket{\tilde{\phi}_\mathrm{GS} (\alpha) | \tilde{\psi}_\mathrm{GS} (\alpha)} = 1$.
Neglecting $O([\hat{H} - \hat{\tilde{H}} (\alpha)]^2)$, we obtain a variational ground-state energy as $\tilde{E}_\mathrm{GS} (\alpha) := \braket{\tilde{\phi}_\mathrm{GS} (\alpha) | \hat{H} | \tilde{\psi}_\mathrm{GS} (\alpha)}$.
Following the principle of minimal sensitivity~\cite{Stevenson1981}, we optimize the variational parameters as $\alpha^* := \arg \min_\alpha \tilde{E}_\mathrm{GS} (\alpha)$.
We regard $E_\mathrm{GS}^\mathrm{MF} := \tilde{E}_\mathrm{GS} (\alpha^*)$ and $\ket{\psi_\mathrm{GS}^\mathrm{MF}} := \ket{\tilde{\psi}_\mathrm{GS} (\alpha^*)}$ as the ground-state energy and eigenvector at the mean-field level, respectively.
Since $\tilde{E}_\mathrm{GS} (\alpha) \geq E_\mathrm{GS}$ should hold if $\hat{H}$ is Hermitian, the method proposed here is an extrapolation of the standard mean-field theory based on the variational principle~\cite{Kleinert2009} to the non-Hermitian regime.

\subsection{Single-site mean-field theory}
\label{Appsub:single_site_mft}

As an application of the method explained above, we consider a single-site mean-field theory for model~\eqref{Eq:model_HCB1lane} though the $\varepsilon$-dependence is not incorporated at this level of theory, as demonstrated below.
We divide the Hamiltonian as $\hat{H} = \hat{\tilde{H}} (\alpha) + [\hat{H} - \hat{\tilde{H}} (\alpha)]$ and take $\hat{\tilde{H}} (\alpha) := \sum_{i = 1}^L \hat{\tilde{H}}_i (\alpha)$ with the single-site Hamiltonian, $\hat{\tilde{H}}_{i} (\alpha)$, defined as
\begin{equation}
    \hat{\tilde{H}}_{i} (\alpha) := - h \sum_{s = \pm} \hat{a}_{i, -s}^\dag \hat{a}_{i, s} - \mu \hat{n}_i - \nu \hat{m}_i^z - \sum_{s = \pm} \lambda_s (\hat{a}_{i, s} + \hat{a}_{i, s}^\dag),
    \label{Eqapp:mean_field_singlesite_hamiltonian}
\end{equation}
Here, $\alpha := \{ \nu, \lambda_+, \lambda_- \}$ is a set of variational parameters: the ferromagnetic mean field $\nu$ and superfluid mean field $\lambda_s$ for particles with spin $s$.
The chemical potential $\mu$ is also introduced to keep the total particle number to $N$ on average.

Writing the ground-state eigenvector and left eigenvector of $\hat{\tilde{H}}_i (\alpha)$ as $\ket{\tilde{\psi}_{\mathrm{GS}, i} (\alpha)}$ and $\ket{\tilde{\phi}_{\mathrm{GS}, i} (\alpha)}$, respectively, we obtain the corresponding eigenvectors for $\hat{\tilde{H}} (\alpha)$ as tensor products:
\begin{equation}
    \left\{
    \begin{array}{l}
        \displaystyle \ket{\tilde{\psi}_\mathrm{GS} (\alpha)} = \bigotimes_{i = 1}^L \ket{\tilde{\psi}_{\mathrm{GS}, i} (\alpha)} \\
        \displaystyle \ket{\tilde{\phi}_\mathrm{GS} (\alpha)} = \bigotimes_{i = 1}^L \ket{\tilde{\phi}_{\mathrm{GS}, i} (\alpha)}.
    \end{array}
    \right.
    \label{Eqapp:mean_field_gs}
\end{equation}
Then, following the method explained above, we can calculate the variational ground-state energy as $\tilde{E}_\mathrm{GS} (\alpha) = \braket{\tilde{\phi}_\mathrm{GS} (\alpha) | \hat{H} | \tilde{\psi}_\mathrm{GS} (\alpha)}$ and obtain the optimized $\alpha$ by minimizing $\tilde{E}_\mathrm{GS} (\alpha)$.
Here, instead of proceeding with the calculation, we focus on the $\varepsilon$-dependence of the ground state.
The $\varepsilon$-dependence of $\tilde{E}_\mathrm{GS} (\alpha)$, $\ket{\tilde{\psi}_\mathrm{GS} (\alpha)}$, or $\ket{\tilde{\phi}_\mathrm{GS} (\alpha)}$ is potentially derived from the following terms in $\braket{\tilde{\phi}_\mathrm{GS} (\alpha) | \hat{H} | \tilde{\psi}_\mathrm{GS} (\alpha)}$:
\begin{equation}
    -\varepsilon t \sum_{i = 1}^L \sum_{s = \pm} s \braket{\tilde{\phi}_\mathrm{GS} (\alpha)  | \hat{a}_{i + 1, s}^\dag \hat{a}_{i, s} - \hat{a}_{i, s}^\dag \hat{a}_{i + 1, s} | \tilde{\psi}_\mathrm{GS} (\alpha)}.
    \label{Eqapp:mean_field_epsilondep}
\end{equation}
However, according to the product forms of $\ket{\tilde{\psi}_\mathrm{GS} (\alpha)}$ and $\ket{\tilde{\phi}_\mathrm{GS} (\alpha)}$ [Eq.~\eqref{Eqapp:mean_field_gs}], $\braket{\tilde{\phi}_\mathrm{GS} (\alpha)  | \hat{a}_{i, s}^\dag \hat{a}_{j, s} | \tilde{\psi}_\mathrm{GS} (\alpha)}$ is reduced to $\braket{\tilde{\phi}_{\mathrm{GS}, 1} (\alpha) | \hat{a}_{1, s}^\dag | \tilde{\psi}_{\mathrm{GS}, 1} (\alpha)} \braket{\tilde{\phi}_{\mathrm{GS}, 1} (\alpha) | \hat{a}_{1, s} | \tilde{\psi}_{\mathrm{GS}, 1} (\alpha)}$ as long as $i \neq j$, and thus Eq.~\eqref{Eqapp:mean_field_epsilondep} is zero.
Consequently, in the single-site mean-field theory, any $\varepsilon$-dependence does not appear in the ground state, and the activity-induced ferromagnetic transition observed in the numerical study (Sec.~\ref{Sec:numerical_results}) is not reproduced.
This is natural since the two-particle bound state explained in Sec.~\ref{Subsec:two_particle_nonmag} is not taken into account in the single-site Hamiltonian~\eqref{Eqapp:mean_field_singlesite_hamiltonian}.
In Sec.~\ref{Sec:mean_field_theory}, we consider a two-site mean-field theory as a minimal self-consistent description of the $\varepsilon$-induced ferromagnetic transition.

\bibliography{ref.bib}

\end{document}